\documentclass[fleqn,usenatbib]{mnras}

\usepackage{newtxtext,newtxmath}

\usepackage[T1]{fontenc}

\DeclareRobustCommand{\VAN}[3]{#2}
\let\VANthebibliography\thebibliography
\def\thebibliography{\DeclareRobustCommand{\VAN}[3]{##3}\VANthebibliography}

\usepackage{graphicx}	
\usepackage{amsmath}	

\title[A metallicity sweet spot for disc fragmentation and planet formation]
{A metallicity sweet spot for disc fragmentation and planet formation}

\author[E. J. Carter et al.]{
Ethan J. Carter,$^{1}$\thanks{E-mail: ecarter6@lancashire.ac.uk}
Dimitris Stamatellos,$^{1}$\thanks{E-mail: dstamatellos@lancashire.ac.uk}, George Blaylock-Squibbs$^{1}$, Alison Young$^{2}$, Ken Rice$^{3,4}$
\\
$^{1}$Jeremial Horrocks Institute for Mathematics, Physics \& Astronomy University of Lancashire, Preston, PR1 2HE\\
$^{2}$School of Physics and Astronomy, University of Leeds, Sir William Henry Bragg Building, Leeds LS2 9JT, UK\\
$^{3}$SUPA, Institute for Astronomy, University of Edinburgh, The Royal Observatory, Blackford Hill, Edinburgh, EH9 3HJ, UK\\
$^{4}$Centre for Planetary Sciences, University of Edinburgh, Edinburgh, EH9 3HJ, UK\\
}

\date{Accepted 2026 August 18. Received 2026 August 18; in original form 2026 May 25}

\pubyear{\the\year{}}

\begin{document}
\label{firstpage}
\pagerange{\pageref{firstpage}--\pageref{lastpage}}
\maketitle

\begin{abstract}
 Fragmentation of gravitationally unstable discs offers an alternate formation mechanism for gas giant planets and brown dwarfs on wide orbits. Metallicity plays a key role in disc evolution from the onset of gravitational instability to the formation of planets. We aim to determine the effect of metallicity on disc fragmentation and on the properties of disc-instability planets. We model gravitationally unstable discs with varying metallicity ($0.01-10\,\rm{Z_{\odot}}$) using the Smoothed Particle Hydrodynamics code {\sc phantom}. 
 Our simulations reveal a “sweet spot” for fragmentation at $0.3\,\rm{Z_{\odot}}$, where cooling is most efficient, with fragmentation also happening less vigorously at higher and lower metallicities. However, further away from the sweet spot, fragmentation becomes more difficult and is eventually suppressed at extreme low and high metallicities ($0.01\,\rm{Z_{\odot}}$ and $10\,\rm{Z_{\odot}}$, respectively), where the disc cools inefficiently. Discs with metallicities close to the sweet spot form more planets per disc, faster, and with lower initial masses than fragmenting discs with higher or lower metallicities. Our results may explain the slight overabundance of wide-orbit giant planets observed around metal-poor stars; these planets may have formed via disc fragmentation.
\end{abstract}




\section{Introduction}

The properties of exoplanets are remarkably diverse. Since the discovery of the first exoplanet orbiting a main-sequence star, 51 Pegasi b \citep{Mayor:1995}, more than 6000 exoplanets have been detected using indirect methods such as radial velocity, transit photometry, and gravitational microlensing, as well as through direct imaging \citep{exoplanet.eu}.

 Most giant planets are thought to form by 
 core accretion, where a rocky or icy core forms via the coagulation of dust particles into larger aggregates that  can then accrete a gaseous envelope after reaching a critical mass \citep{Mizuno:1980,Stevenson:1982,Pollack:1996b}. However, the core accretion scenario struggles to explain the formation of giant planets at distances greater than $20\rm \,AU$. Furthermore, the formation of giant planets in the core accretion scenario requires a timescale on the order of a few Myr which may exceed the expected lifetime of the disc, although pebble accretion may accelerate this process, allowing giant planets to form by core accretion on shorter timescales \citep{Lambrechts:2012a}. The existence of giant planets on wide orbits, particularly those around M-dwarfs, also poses a challenge to the core accretion scenario. Additionally, \cite{Johnson:2012} show that core accretion struggles to explain the formation of massive planets orbiting metal-poor stars. They investigate the minimum metallicity required for planet formation through comparing the timescale for dust settling (which precedes planetesimal formation) in the midplane to the expected disc lifetime, and find a relationship between the lower limit for the critical iron abundance for planet formation and distance $r$ from the host star $[\mathrm{Fe/H}]_{\mathrm{crit}}\simeq-1.5+\operatorname{log}(r/1\,\mathrm{AU})$. 
 This indicates that planet formation at large orbital radii is particularly challenging in low-metallicity discs, suggesting that wide-orbit giant planets are unlikely to form at the lowest metallicities within the core accretion framework.

Fragmentation due to gravitational instabilities (GI) in a disc  provides an alternative giant planet formation mechanism. In this scenario, planet formation occurs as a consequence of gravitational fragmentation in young protostellar discs \citep{Kuiper:1951b,Cameron:1978b,Boss:1997,Rice:2003,Stamatellos:2007a, Rice:2022}. A disc becomes gravitationally unstable when it is sufficiently massive so that its own self gravity dominates over thermal and rotational support, i.e. when the Toomre criterion \citep{Toomre:1964} is satisfied,
\begin{equation}
    Q\equiv \frac{c_{s}(R)\Omega(R)}{\pi G\Sigma(R)}\lesssim\,Q_{crit}\simeq1,
\end{equation}
where $Q$ is the Toomre parameter, $c_{s}$ is the sound speed, $\Omega$ is the angular frequency and $\Sigma$ is the surface density of the disc at a given orbital radius $R$. Gravitational instabilities in the disc lead to the formation of spiral arms, transferring angular momentum radially outwards. If the disc can cool fast enough (typically within a few outer orbital periods, i.e. $~\tau_{\mathrm{cool}}\lesssim3\Omega^{-1}$), gravitational instabilities can lead to fragmentation \citep{Gammie:2001b}, forming dense clumps of gas which may collapse further on a dynamical timescale and evolve into a giant planet or brown dwarf \citep{Rice:2003, Stamatellos:2007a, Stamatellos:2009b}. Even lower mass (i.e. super-Earth) planets could also form by GI through tidal downsizing \citep{Nayakshin:2017}. \cite{Deng:2021} also suggest that magnetic fields may reduce the mass of planets formed by GI, leading to the formation of intermediate-mass planets.

The core accretion scenario predicts metal-rich giant planets \citep{Pollack:1996b}, whereas disc-instability planets (i.e. those formed by GI) might be expected to be metal-poor, initially reflecting the metallicity of the host disc. However, disc-instability planets may subsequently become enriched in heavy elements through the accretion of solids after their formation \citep{Helled:2006a}. \cite{Helled:2010} estimate the mass of heavy-element-rich planetesimals that could have been accreted by the massive planets orbiting HR~8799 if these planets formed via GI. Their results suggest that the planets in the HR~8799 system are expected to have metallicities similar to that of their host star.

\begin{figure}
    \centering
    \includegraphics[width=1\linewidth]{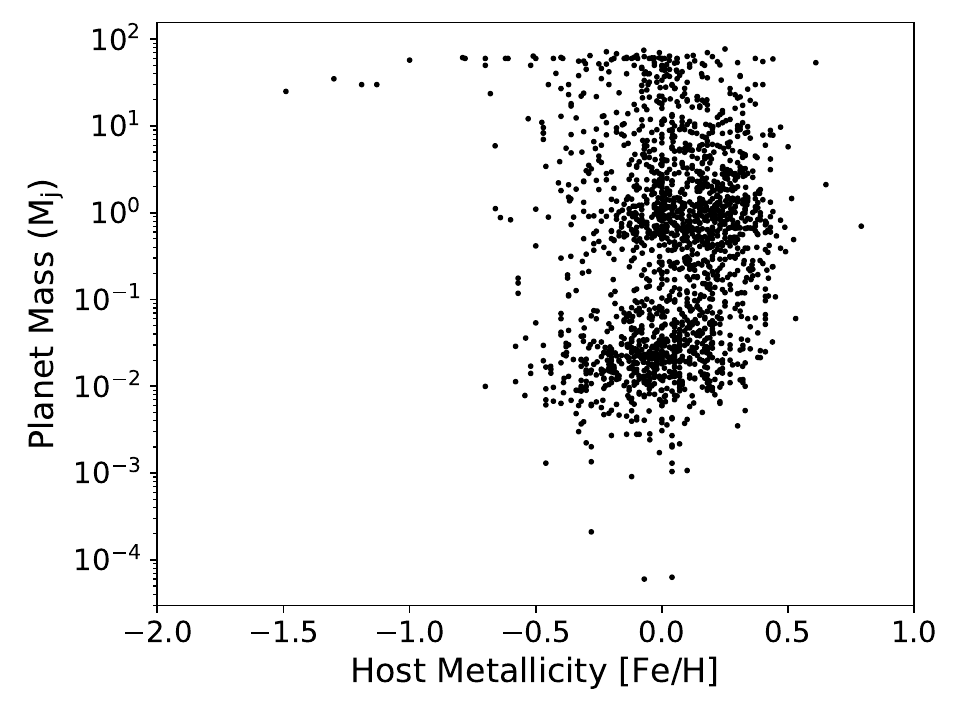}
    \caption{Exoplanet mass plotted against the metallicity  of the host star (data from the exoplanet.eu database; accessed 12/06/2026).}
    \label{fig:obs_mass_metallicity}
\end{figure}

\begin{figure}
    \centering
    \includegraphics[width=1\linewidth]{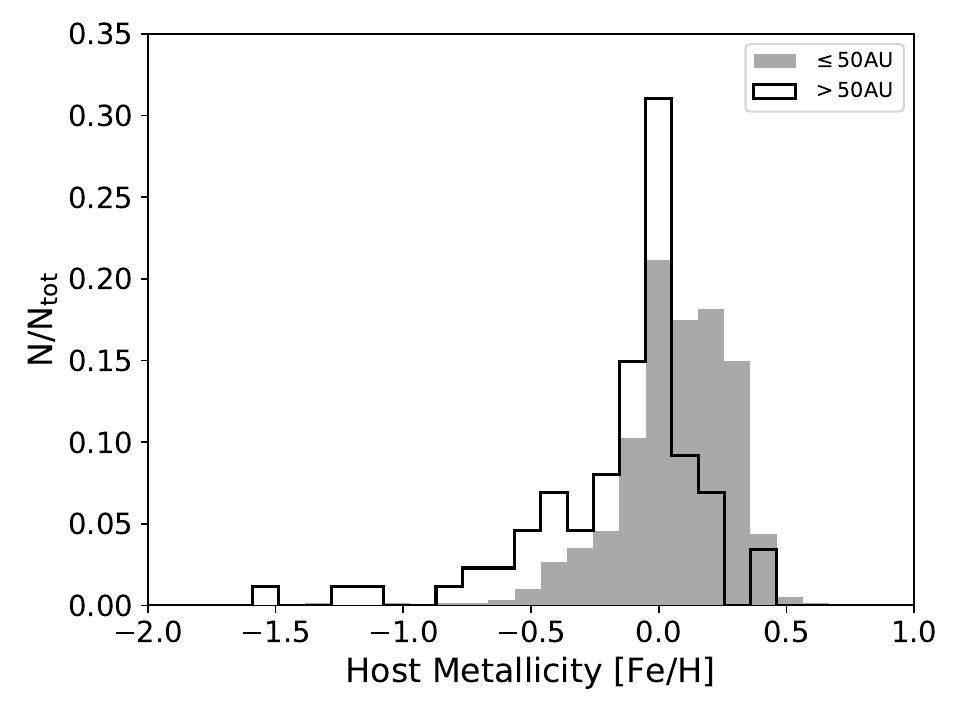}
    \caption{The metallicity distributions of host stars harbouring confirmed massive (${M_{\rm p}}\geq1\,\mathrm{M}_{\rm J}$) exoplanets, separated into systems with semi-major axes greater and smaller than $50\rm\,AU$  (data from the exoplanet.eu database; accessed 12/06/2026).}
    \label{fig:obs_metallicity_hists_jups}
\end{figure}

Exoplanets are found around stars with a range of metallicities; mostly from $ {\rm [Fe/H]=-0.5}$ to ${\rm [Fe/H]=0.5}$, i.e. from $\sim 0.3$ to $\sim 3{\rm \, Z}_{\odot}$ (see Fig. \ref{fig:obs_mass_metallicity}). However, the probability of finding a gas giant planet around a star increases with the host star's metallicity \citep{Fischer:2005a, Howe:2026a}. Planet mass typically increases with stellar metallicity up to $\sim 1\,\mathrm{Z}_{\odot}$ (see Fig. \ref{fig:obs_metallicity_hists_jups}) \citep{Buchhave:2014, Santos:2017, Narang:2018}. This trend is in support of the core accretion scenario, which is more likely in a solid-rich, high-metallicity environment where a large core may form more easily \citep{Matsukoba:2023}. However, this positive correlation between stellar metallicity and planet mass appears to invert for planets with masses $\geq 4\rm\, M_{J}$, where planet mass instead increases as stellar metallicity decreases \citep{Santos:2017, Narang:2018, Schlaufman:2018, Maldonado:2019}. \cite{Swastik:2021} show that for directly imaged planets (massive planets on wide orbits, $\geq 10\rm~\,AU$), this negative correlation between stellar metallicity and planet mass for super-Jupiters is stronger, suggesting that Jupiter-like and lower-mass planets are likely to form via core accretion, whilst super-Jupiters and brown dwarfs are likely to form via fragmentation as a result of gravitational instabilities in the disc. 
Fig. \ref{fig:obs_metallicity_hists_jups} shows the distribution of metallicities of the host stars of observed exoplanets with orbital radii above and below $50~\rm{\,AU}$. Giant exoplanets ($\geq1\,\rm{M_{j}}$) with orbital radii $\geq50\,\rm{AU}$ are more commonly observed orbiting stars with lower metallicities compared to exoplanets with orbital radii $\leq50\,\rm{AU}$. This suggests that close-in and wide-orbit giant planets may form through different mechanisms.

Numerical and analytical studies have considered the effect of metallicity on star and planet formation. \cite{Boss:2002} find that the gravitational instability (and subsequent planet formation) develops similarly for discs with metallicities $0.1$, 1 $10\rm\,Z_{\odot}$, suggesting that disc fragmentation is independent of metallicity. However, \cite{Cai:2006} found that GI is in fact sensitive to metallicity within the range $0.25\rm\,Z_{\odot} \leq Z \leq 2\rm\,Z_{\odot}$, with the GI becoming stronger for lower metallicity; whilst discs with subsolar metallicities were gravitationally unstable, fragmentation did not occur.  \cite{Lee:2025} find that fragmentation is promoted in low-opacity discs, which effectively corresponds to low-metallicity environments. Other studies show that whilst the minimum disc fragmentation mass may be insensitive to metallicity, it may still have an effect on the disc evolution \citep{Mercer:2020, Matsukoba:2022, Matsukoba:2023}. In metal-poor discs, \cite{Mercer:2020} find that the disc may take longer to fragment due to the reduced opacity and metallicity, though fragmentation is more rigorous when it does occur. They also find that in high-metallicity discs, fragmentation may be suppressed as the disc cannot cool fast enough; instead the disc expands and achieves a self-regulated state, with $Q\sim1$.

In the context of star formation, studies show that whilst metallicity does not play a role in determining the properties of stellar systems, lower metallicity leads to a delayed onset of star formation due to less efficient cooling (though the star formation rate remains similar irrespective of metallicity at later stages) \citep{Bate:2014, Bate:2019}. It has also been shown that more massive substellar companions are expected to form in metal-poor clouds, leading to a larger population of brown dwarfs and more  mergers than in metal-rich environments \citep{Tanaka:2014, Bate:2014, Maldonado:2019, Bate:2019}.

The aim of this paper is to investigate the role of metallicity in disc evolution and planet formation via GI. We model the evolution of gravitationally unstable discs with varying metallicity and examine (i) whether such discs fragment, and (ii) the properties of the resulting protoplanets. We use the term {\it protoplanets} to refer to objects formed via fragmentation in protostellar discs, irrespective of their mass; some may end up as planets but others may grow further in mass to become brown dwarfs. We define disc fragmentation, and hence protoplanet formation, to occur when the central density of a clump that forms in the disc  reaches $10^{-3}\,\mathrm{g\,cm^{-3}}$.

In section 2, we discuss the computational method  used and the set-up of the simulations. In section 3, we discuss the effect of metallicity on the evolution on gravitationally unstable discs and fragmentation. In section 4, we present our results regarding the effect of metallicity on the properties of disc-instability protoplanets and in section 5, we discuss our results regarding the importance of metallicity on fragmentation in gravitationally unstable discs.  


\section{Computational Method \& System Setup} 
We model the thermodynamics of massive, gravitationally unstable discs using the three-dimensional Smoothed Particle Hydrodynamics (SPH) code {\sc phantom} \citep{Price:2018}. {\sc phantom} is based on the SPH technique, independently developed by \citet{Lucy:1977} and \citet{Gingold:1977}, which solves the equations of hydrodynamics in Lagrangian form.

\subsection{Disc Thermodynamics}

We model the radiative transfer processes which regulate the heating and cooling in the disc with the approximation method of \cite{Lombardi:2015}, a modified approach of the method from \cite{Stamatellos:2007a}. The method uses an SPH particle's density $\rho$, temperature $T$ and the pressure scale-height $H_{P}\equiv\,P/|\Delta\,P|$ to estimate the mean Rosseland and Planck optical depths ${\tau_R}$ and ${\tau_P}$ for the particle. It then uses them to determine the radiative heating/cooling rate,
\begin{equation}
\label{eq:optical_depth}
\frac{du}{dt}
=
\frac{4\sigma_{\rm SB}(T_{\rm 0}^4-T^4)}
{{\bar \Sigma}\left(\tau_R + \tau_P^{-1}\right)},
\end{equation}
where $\sigma_{S_{B}}$ is the Stefan-Boltzmann constant, $T_{0}$ is the background temperature below which the gas particle cannot cool radiatively (set by the stellar heating). 
$\bar{\Sigma}$ is the mass-weighted mean mass column density and is calculated using the pressure scale-height, which is a good approximation in discs,
\begin{equation}
    \bar{\Sigma} = \zeta'\frac{P}{|\textbf{a}_{h}|},
\end{equation}
where $P$ is the gas pressure and $\zeta'$ is a dimensionless coefficient ($\zeta'=1.014$). The term $\textbf{a}_{h}$ is the hydrodynamical acceleration and is given by
\begin{equation}
    \textbf{a}_{h}=\frac{-\nabla P}{\rho}.
\end{equation}
Using
\begin{equation}
\tau_{\rm R}={\bar \kappa}_{\rm R}(\rho,T){\bar\Sigma},
\qquad
\tau_{\rm P}={\kappa_{\rm P}}(\rho,T){\bar\Sigma},
\end{equation}
we have
\begin{equation}
    \label{eq:dudt_rad}
    \frac{du}{dt}
    =\frac{4\sigma_{S_{\mathrm{B}}}(T_{0}^{4}(r)-T^{4})}{\bar{\Sigma}^{2}\bar{\kappa}_{\mathrm{R}}(\rho,T)+\kappa_{\mathrm{p}}^{-1}(\rho,T)}, 
\end{equation}
where $\bar{\kappa}_{\mathrm{R}}$ is the Rosseland-mean opacity, and $\kappa_{\mathrm{p}}$ is the Planck-mean opacity (we set  $\bar{\kappa}_{\mathrm{R}}\approx\kappa_{\mathrm{p}}$).
The \cite{Lombardi:2015} method provides a better approximation of the radiative heating and cooling in protoplanetary discs, and a more efficient cooling than the \cite{Stamatellos:2007a} method \citep{Mercer:2018,Young:2024}, and hence is expected to allow discs to fragment at lower masses \citep{Teasdale:2026b}.

\subsubsection{Opacity}

We use the \cite{Bell:1994} parameterisation of  opacities,
\begin{equation}
    \kappa\left( \rho,T \right)=\kappa_{0}\rho^{a}T^{b},
\end{equation}
where $\left( \kappa_{0}, a,  b \right)$ are constants depending on the dominant processes contributing to the opacity in different temperature and density regimes (see Table \ref{tab:opacities}). This parameterisation accounts for the effects of ice melting, dust sublimation, molecules, $\rm H^{-}$ absorption, bound-free transitions, free-free transitions and electron scattering. 

To vary the metallicity, we scale the dust opacity by a factor $z$, such that $z=1$ corresponds to solar metallicity. We also scale the gas molecular opacity by the same factor, as a higher abundance of dust grains leads to an increased abundance of molecules. We do not modify the gas opacity where $\rm H^{-}$ absorption, bound-free transitions, free-free transitions and electron scattering are the dominant processes. The range over which the factor $z$ is applied is determined by the regime in which molecular opacity dominates over $\rm H^{-}$ absorption, i.e.
\begin{equation}
    10^{-8}\,\rho^{2/3} T^{3} \geq 
    10^{-36}\,\rho^{1/3} T^{10}\,.
    \label{eq:opac_regime}
\end{equation}

We assume that dust is coupled with the gas and we do not model the dynamics of dust nor the effects of dust evolution in our simulations. We neglect processes such as vertical settling of dust towards the midplane, grain growth, and radial drift which serve to redistribute the dust in the disc, affecting the local opacity and the cooling of the disc.
Furthermore, we do not include additional cooling processes, such as line emission from H$_2$ and HD molecules or fine-structure line emission from O{\sc i} and C{\sc ii}, which may enhance cooling at low metallicities \citep{Matsukoba:2022,Matsukoba:2023}.


\begin{table}
\centering
\begin{tabular}{lccc}
\hline
Dominant process            & $\kappa_{0}   $   & $a  $ & $b   $  \\ \hline
Ice grains                  & $2$x$10^{-4}   $  & $0  $ & $2   $  \\
Evaporation of ice grains   & $2$x$10^{16}   $  & $0  $ & $-7  $  \\
Metal grains                & $0.1           $  & $0  $ & $1/2 $  \\
Evaporation of metal grains & $2$x$10^{-81}  $  & $1  $ & $-24 $  \\
Molecules                   & $10^{-8}       $  & $2/3$ & $3   $  \\
H- absorption               & $10^{-36}      $  & $1/3$ & $10  $  \\
bf and ff transitions       & $1.5$x$10^{-20}$  & $1  $ & $-5/2$  \\
Electron scattering         & $0.348         $  & $0  $ & $0   $  \\ \hline
\end{tabular}
\caption{Opacity law parameters from \protect\cite{Bell:1994}}
\label{tab:opacities}
\end{table}

\subsection{Initial Conditions}
We model a disc of mass $ M_{D}=0.25 \rm\, M_{\odot}$ hosted by a star with mass $M_{*}=0.7 \rm\, M_{\odot}$. This corresponds to a disc-to-stellar mass ratio of $0.36$, chosen such that a $1\,\rm{Z_{\odot}}$ disc is just beyond the threshold of fragmentation. The disc metallicity is set to  $Z=0.1,0.3,1,3, {\rm and}\ 10\,\rm{Z_{\odot}}$. The disc extends from $1-120\rm\,AU$ and is represented by $5\rm\,x10^{5}$ SPH particles. The surface density profile of the disc $\Sigma$ is given by
\begin{equation}
    \label{eq:sigma}
    \Sigma=\Sigma_{0}\left(\frac{R}{\rm\,1AU}\right)^{-1}\,\,\left(1-\sqrt{\frac{\rm\,1AU}{R}}\right),
\end{equation}
where $\Sigma_{0}=3.5\rm\times10^{3}\,g\,cm^{-2}$, $R$ is the radial distance to the host star. We set the disc temperature profile of the disc using
\begin{equation}
    \label{eq:temp}
    T(R)={T_\mathrm{1AU}}\left(\frac{R}{\mathrm{1AU}} \right)^{-0.6},
\end{equation}
where $\,{T}_\mathrm{1AU}=250\mathrm{K}$. ${T}(R)$ is also used to set the background temperature, $T_{0}$, in Eq. \ref{eq:dudt_rad}.
We use the disc viscosity prescription as described in \cite{Lodato:2010, Price:2018} which describes viscosity using an effective Shakura-Sunayev shear viscosity with parameters $\alpha_{\rm ss}=0.005$, $\alpha_{\rm SPH}=0.1$ and $\beta_{\rm SPH}=2.0$. We simulate multiple realisations of the same initial disc parameters per metallicity, where for each realisation the disc is initialised independently with a different random seed which determines the initial positions of particles in the disc.

\subsection{Sink Particles}

The star that hosts the disc is modelled as a sink particle, which only interacts with the rest of the computational domain through its gravity \citep{Bate:1995}. We assign the sink accretion radius of the star as $ \mathrm{R_{sink,*}}=1$\,AU. A protoplanet, also represented by a sink particle, forms when the density reaches $\rho_{c}=10^{-3}\rm\,g\,cm^{-3}$. Sink particles that represent protoplanets have an accretion radius of $R_{\rm sink,p}=0.1$\,AU. Gas particles within the sink accretion radius are accreted onto the sink particles unconditionally. This allows us to follow the collapse of a condensation through the first and second hydrostatic core phases \citep{Stamatellos:2009a, Fenton:2024}. Sink particles are allowed to merge if they are within a separation $R\leq0.4$\,AU and are bound. Sinks are forced to merge unconditionally if they are within $R\leq0.2$\,AU.


\section{The effect of metallicity on disc evolution and fragmentation}
\begin{figure*}
    \centering
    \includegraphics[width=\textwidth]{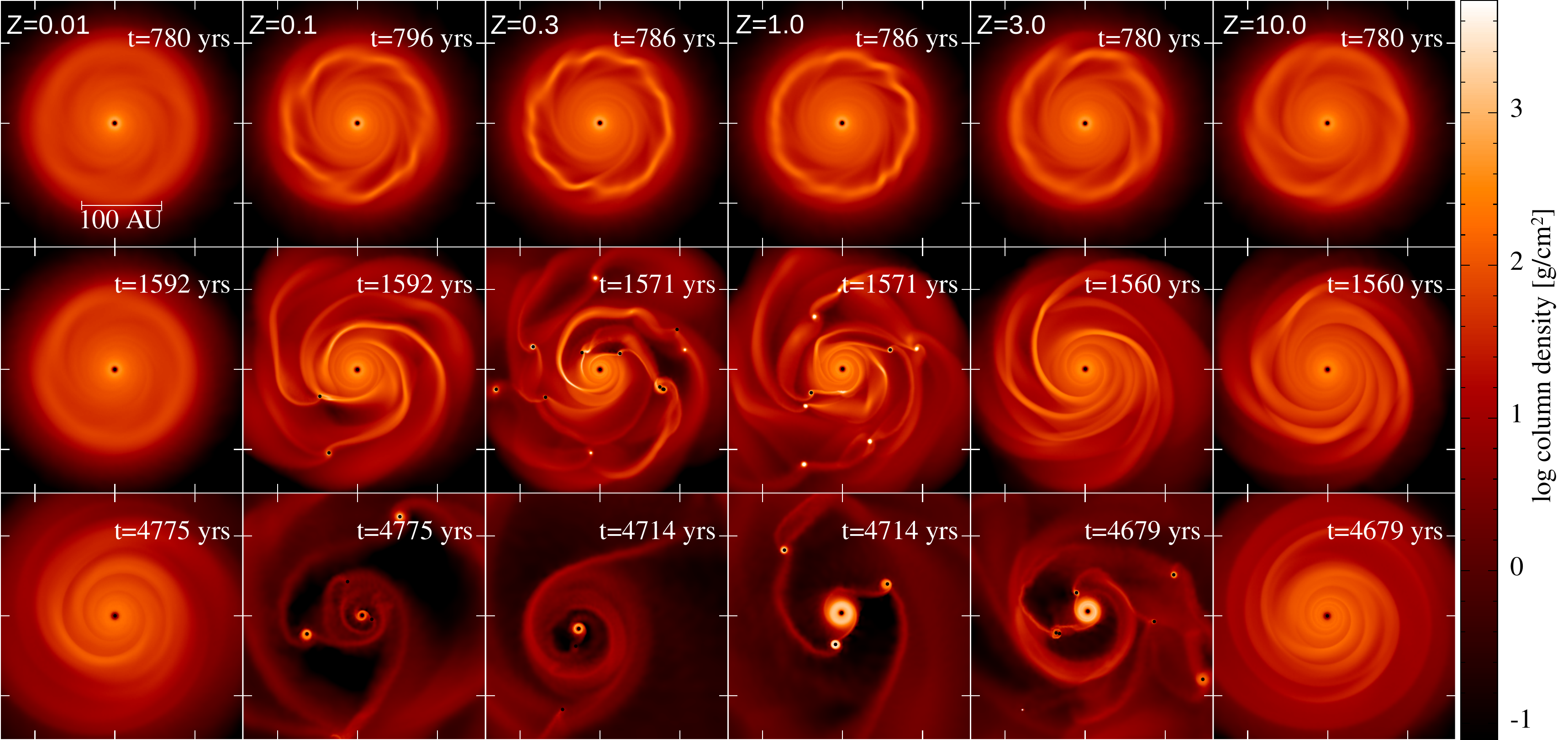}
    \caption{Evolution of the surface density (top to bottom) for a representative disc at different metallicities (left to right). The initial conditions are such that the disc is gravitationally unstable in all runs, which results in the formation of spiral arms. However, fragmentation is suppressed in all simulations with $ Z=[0.01,10]\,\rm{Z}_{\odot}$, and in $25\%$ of the simulations with $ Z = 3\,\rm{Z}_{\odot}$.}
    \label{fig:disc_evolution}
\end{figure*}

We simulate the evolution of gravitationally unstable discs with varying metallicity ($ Z=0.01,0.1,0.3,1,3,10\,\rm{Z_{\odot}}$). We show a typical outcome of representative simulation for each metallicity in Fig.~\ref{fig:disc_evolution}. The disc is gravitationally unstable and develops strong spiral arms on a relatively short timescale ($\leq 1\,\rm kyr$). As the disc continues to evolve the disc fragments in all simulations with metallicities $ Z=0.1,0.3, {\rm and}\ 1.0\,\rm{Z_{\odot}}$, forming dense condensations of gas which collapse on a dynamical timescale and become protoplanets, in agreement with prior studies \citep{Stamatellos:2009b,Mercer:2020}. Fragmentation is suppressed in $25\%$ of simulations of discs with $ Z=3.0\,\rm{Z_{\odot}}$, and is suppressed in all simulations in the highest and lowest metallicity models ($ Z=0.01,10\,\rm{Z_{\odot}}$). After protoplanets have formed, they proceed to interact gravitationally through scattering events as they continue to grow and accrete material from the disc. Mergers between protoplanets are also quite common \citep{Boss:2023}. We follow each simulation until the disc has lost $70\%$ of its original mass (typically around $\sim4-5$ kyr); this mass has been accreted either onto the protoplanets or onto the central star. We examine the protoplanet properties across all simulations at this threshold so that we can compare the protoplanet properties at the same evolutionary stage of each disc.

\begin{table*}
    \begin{tabular}{ccrrrrrr}
    \hline
    $Z ({\rm Z}_{\odot})$ & Runs & $ N_{p}$ & $\left\langle \mathrm{N_{\rm p}} \right\rangle$ & ${N_{\rm merged}} (\%)$ & ${N_{\rm final}} (\%)$ & ${N_{\leq20\rm{M_{\rm J}}}} (\%)$ & ${N_{>20\rm{M_{\rm J}}}} (\%)$ \\
    \hline
    0.1 & 10 & 115 & 12$\pm$2 & 37$\pm$6 &  63$\pm$7 & 54$\pm$9 & 46$\pm$9 \\
    0.3 & 10 & 143 & 14$\pm$2 & 36$\pm$5 &  64$\pm$7 & 70$\pm$9 & 30$\pm$6 \\
    1.0 & 10 & 106 & 10$\pm$1 & 33$\pm$6 &  67$\pm$8 & 62$\pm$9 & 38$\pm$7 \\
    3.0 & 20 & 78  &  4$\pm$1 & 30$\pm$6 &  70$\pm$10 & 45$\pm$9 & 55$\pm$10 \\
    \hline
    \end{tabular}
    \caption{General statistics for each set of simulations at different metallicities $Z$. We present the number of simulations per metallicity (Runs), the total number of protoplanets formed, $ N_{p}$,  the mean number of protoplanets formed per disc, $\left\langle \mathrm{N_{p}}\right\rangle$, the percentage of protoplanets that have merged, $ N_{\rm merged}$, the percentage of protoplanets that survived until the end of the simulation,$ N_{\rm final}$, and the percentage of those with final mass above and below  $20\rm{M_{J}}$. The errors on each value are calculated assuming Poisson statistics. We performed twice as many simulations with $Z=3\,\rm{Z}_{\odot}$ in order to produce a number of planets that is comparable with the that of discs with the other metallicities (as only 25\% of the  $Z=3\,\rm{Z}_{\odot}$ discs fragment). Discs with $Z=0.01,10\,\rm{Z}_{\odot}$ do not fragment.}
        \label{tab:stats}
    \end{table*}


We use the method of \cite{Sleath:1996} to quantify the strength of the spiral arms in a typical simulation for each metallicity by calculating their amplitude. The structures in the disc may be decomposed into a sum of Fourier components using logarithmic spirals as a basis. A logarithmic spiral may be described by
\begin{equation}
    R=R_{0}\rm e^{-m\phi/\zeta},
\end{equation}
where $R$ is the radial position, $\phi$ is the azimuthal angle of the particle, $m$ is the mode of the perturbation and $\zeta=-m/\tan(\beta)$ represents the pitch angle of the spiral, $\beta$. 
The Fourier transform is expressed by \cite{Sleath:1996} as
\begin{equation}
    \begin{aligned}
        F(\zeta,m)=\int_{-\infty}^{+\infty} \int_{-\pi}^{+\pi}\sum^{N}_{j=1}\{\delta(u-\operatorname{ln}[R_{j}]\delta(\phi-\phi_{j})\}
        &\\\\\ \times e^{-i(\zeta u+m\phi)}dud\phi=\frac{1}{N}\sum^{N}_{j=1}{e}^{-i(\zeta \operatorname{ln}[R_{j}]+m\phi_{j})},
    \end{aligned}
\end{equation}
where $(R_{j},\phi_{j})$ are the co-ordinates of particle $j$. For the discs that fragment we do not include particles with densities above $10^{-9}\, {\rm g\,cm^{-3}}$ to exclude the fragments themselves and capture only the spiral arms.

\begin{figure}
    \centering
    \includegraphics[width=0.95\linewidth]{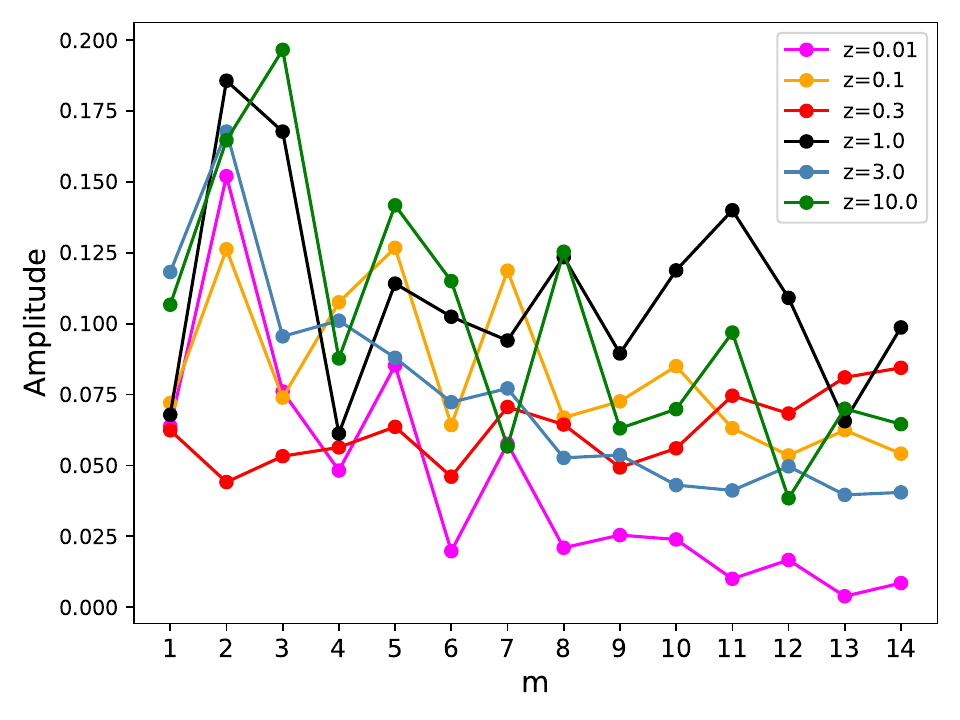}
    \caption{Power spectrum of the spiral modes for a typical simulation just after fragmentation, for each metallicity. To capture only the spiral arms, we exclude fragments from our calculations. The $m = 2$ mode is the dominant mode, but there is not a strong correlation between metallicity and the strength of the spirals arms.}
    \label{fig:fourier}
\end{figure}

Fig. \ref{fig:fourier} shows the power spectrum for a typical disc simulation for each metallicity just after fragmentation.  For simulations in which fragmentation was suppressed, we plot the power spectrum after the spiral arms develop. Generally, the disc retains a dominant $m = 2$ mode
in agreement with previous studies \citep{Forgan:2011,Longarini:2025,Teasdale:2026b}. The power spectrum is largely independent of metallicity. In simulations in which the disc fragments, the spiral arms are disrupted as they become sites for protoplanet formation, leading to a more complex morphology (see lines for ${Z}=0.1,1.0\,\rm{Z_{\odot}}$). Discs with metallicity ${Z}=0.3\,\rm{Z_{\odot}}$ cool most efficiently (see Discussion) and form a larger number of protoplanets within a short timescale, significantly disrupting the disc and preventing the development of strong spiral arms. In discs with metallicity $ Z=0.01,{\rm and}\ 10\,\rm{Z_{\odot}}$, fragmentation is suppressed, as they cannot cool fast enough. We find that discs with the lowest metallicity $ Z=0.01\,\rm{Z_{\odot}}$ typically develop weaker spiral arms than the discs with the highest metallicity $ Z=10\,\rm{Z_{\odot}}$. However, despite this we do not see a strong correlation between metallicity and the strength of spirals arms that form in the disc.

\begin{figure}
    \centering
    \includegraphics[width=0.95\columnwidth]{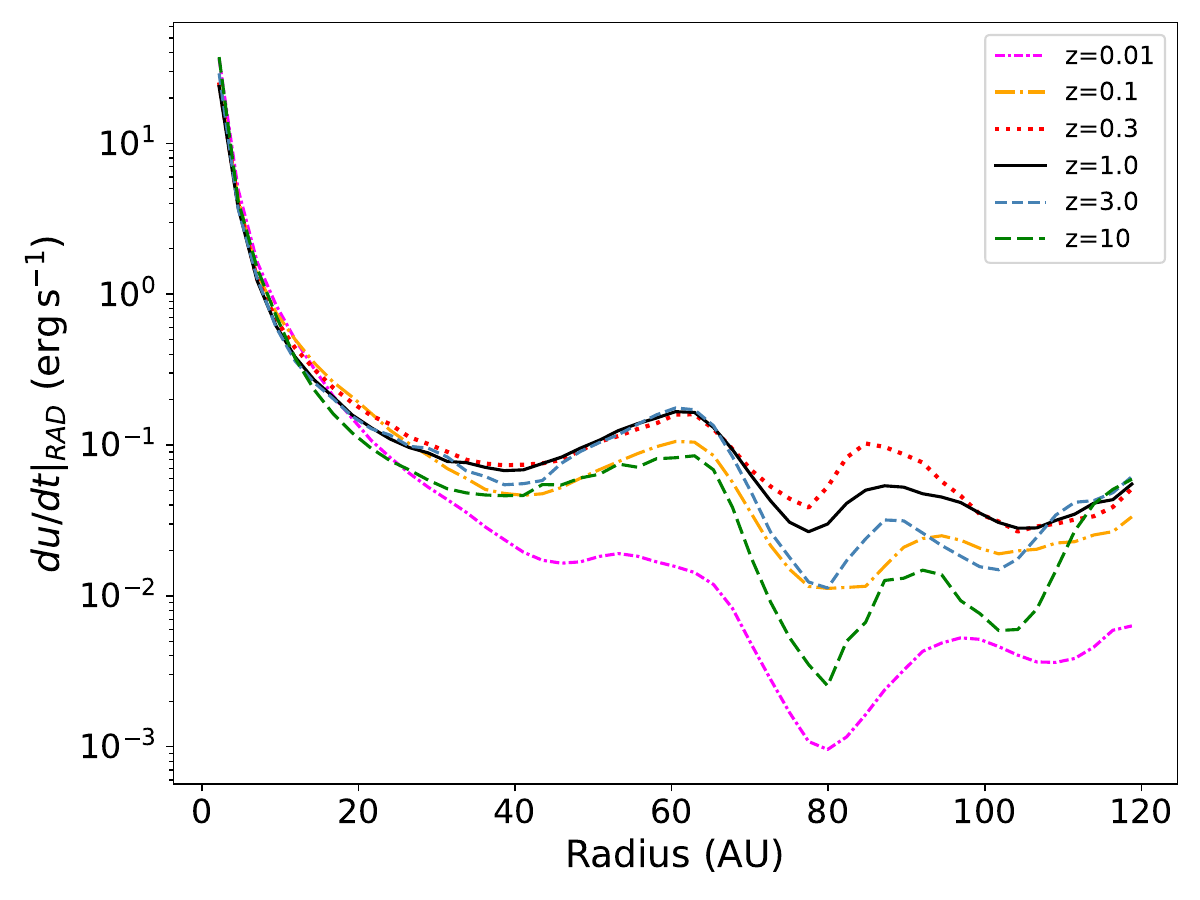}
    \caption{Azimuthally averaged radiative cooling rate $\rm du/dt$ for each metallicity disc against radius at $t=950\,{\rm yr}$ (before fragmentation). Discs with metallicity $0.3\rm\,Z_{\odot}$ cool the most efficiently.}
    \label{fig:dudt_cooling_rates}
\end{figure}

We calculate the azimuthally averaged radiative cooling rate (see Eq. \ref{eq:dudt_rad}), $\rm du/dt$  throughout each disc at $t=950\,{\rm yr}$, i.e. before any of the discs fragment (see Fig. \ref{fig:dudt_cooling_rates}). We find that the cooling rate is similar for all metallicities within $\sim 20\,{\rm AU}$, where the disc temperature is dominated by heating from the star (see Eq. \ref{eq:temp}). Beyond $20 {\rm AU}$, the cooling rate drops in the $\rm 0.01\,\rm{Z_{\odot}}$ (magenta line) and $\rm 10\,\rm{Z_{\odot}}$ (green line) discs as they are extremely optically thin or optically thick, respectively, and hence cool less efficiently. The cooling rate for discs with metallicity $Z=0.3,1.0,3.0\,\rm Z_{\odot}$ (red, black, blue lines) is similar until the $\sim 70-100 \,\rm AU$ region; this is the region where the first protoplanets form in each disc (see Fig. \ref{fig:jups_BDs_formation_radius}.In this region, we find that the $0.3\rm\,Z_{\odot}$ disc cools most efficiently, whilst the $1.0\rm\,Z_{\odot}$ and $3.0\rm\,Z_{\odot}$ cool less efficiently as they are more optically thick. The $0.01\rm\,Z_{\odot}$ disc (orange line) also has a slightly lower cooling rate throughout the disc as it is more optically thin.

\begin{figure}
    \centering
    \includegraphics[width=0.95\linewidth]{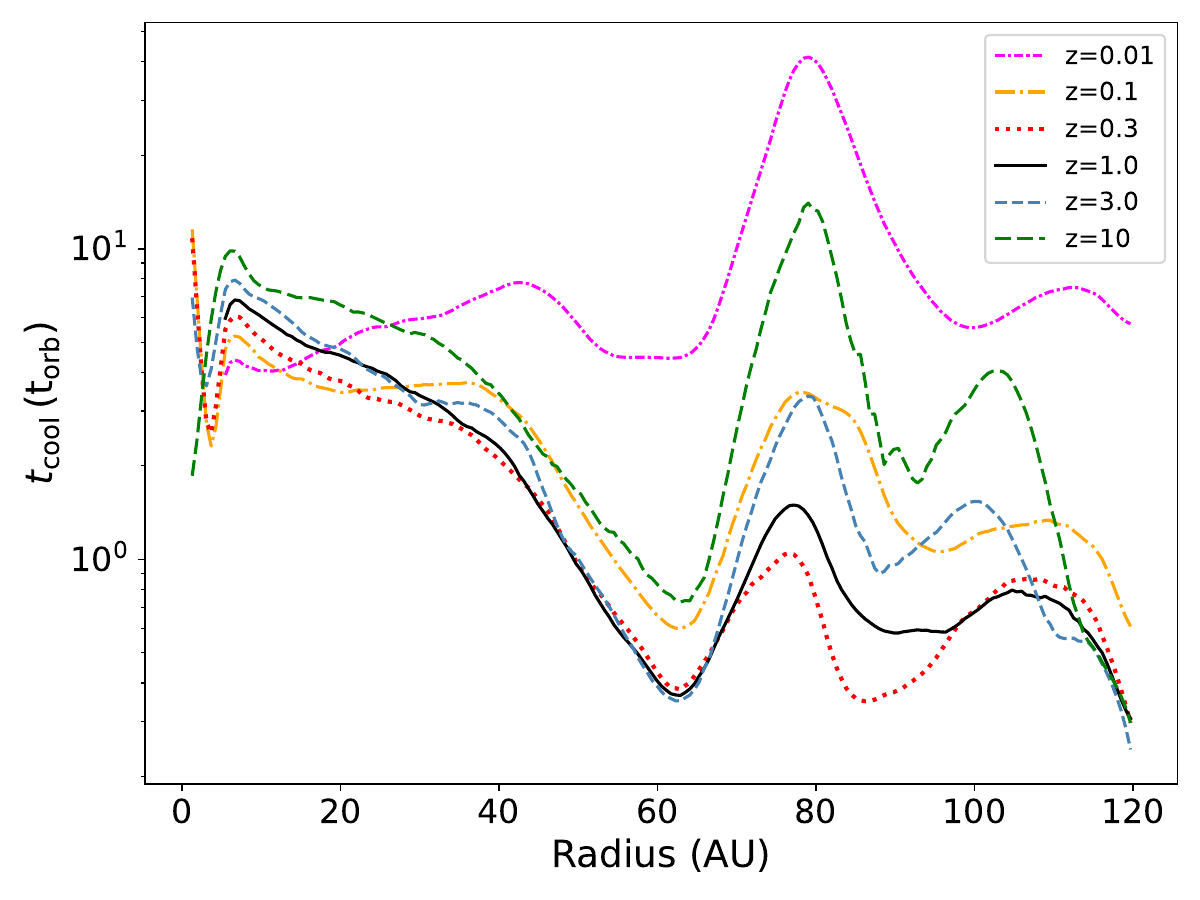}
    \caption{Azimuthally averaged radiative cooling timescale $t_{\rm cool}$ against radius for each metallicity disc  at $t=950\,{\rm yr}$. Discs with metallicity $0.3\rm\,Z_{\odot}$ have the shortest cooling timescales compared to discs with lower or higher metallicity.}
    \label{fig:tcool_profiles}
\end{figure}

We also calculate the radiative cooling timescale, $t_{\rm cool}=du/(du/dt)$. We plot the azimuthally averaged radiative cooling timescale in Fig. \ref{fig:tcool_profiles}. We find that discs with metallicities $Z=1.0$ and $0.3\rm\,Z_{\odot}$ have shorter cooling timescales in the outer regions of the disc compared to discs with lower or higher metallicities. We find that the $0.3\rm\,Z_{\odot}$ disc has a cooling timescale shorter than the orbital period in the $\sim 80-90 \,\rm AU$ region, where the first fragments form.  

The fact that the $0.3\rm\,Z_{\odot}$ disc cools most efficiently compared to simulations with lower and higher metallicities suggests a non-monotonic relationship between metallicity and the growth of the gravitational instability, with a metallicity sweet spot around $0.3\rm\,Z_{\odot}$ (see Discussion).

\section{The effect of metallicity on the properties of disc-instability protoplanets}

In this section we discuss the properties of the disc-instability protoplanets that  form in the disc simulations. In Table~\ref{tab:stats} we present some general statistics for the protoplanets that form in the simulations. We exclude the simulations with $ Z=0.01,10\,\rm{Z_{\odot}}$, as fragmentation is fully suppressed in all realizations and hence no protoplanets form. For simulations with disc metallicities $ Z=0.1,0.3,1.0\,\rm{Z_{\odot}}$ we find a similar mean number of protoplanets formed per disc, but simulations with $ Z=0.3\,\rm{Z_{\odot}}$ form the highest number of protoplanets per disc. Discs with higher metallicity, $ Z=3.0\,\rm{Z_{\odot}}$, form fewer protoplanets per disc. Across all simulations, protoplanets form with a broad range of final masses. Not all protoplanets survive until the end of the simulations; some are accreted onto the host star, while others merge with each other.

 We also present statistics for protoplanets with final masses above and below $20\,\mathrm{M_{J}}$ (see Table~\ref{tab:stats}). We adopt this threshold as objects with masses $\leq 20\,\mathrm{M_{J}}$ may possibly evolve into planets, whereas those with masses $> 20\,\mathrm{M_{J}}$ are more likely to become brown dwarfs. Discs with $Z =  [0.3,1.0]\,\mathrm{Z}_{\odot}$ preferentially form lower-mass protoplanets, while the $Z = 3\,\mathrm{Z}_{\odot}$ discs form slightly more objects with $> 20\,\mathrm{M_{J}}$ than objects with $\leq 20\,\mathrm{M_{J}}$. The $Z = 0.3\,\mathrm{Z}_{\odot}$ disc yields the highest fraction of low-mass protoplanets.


\subsection{Protoplanet mass, orbital radius, and formation time}

\begin{figure}
    \centering
    \includegraphics[width=0.9\columnwidth]{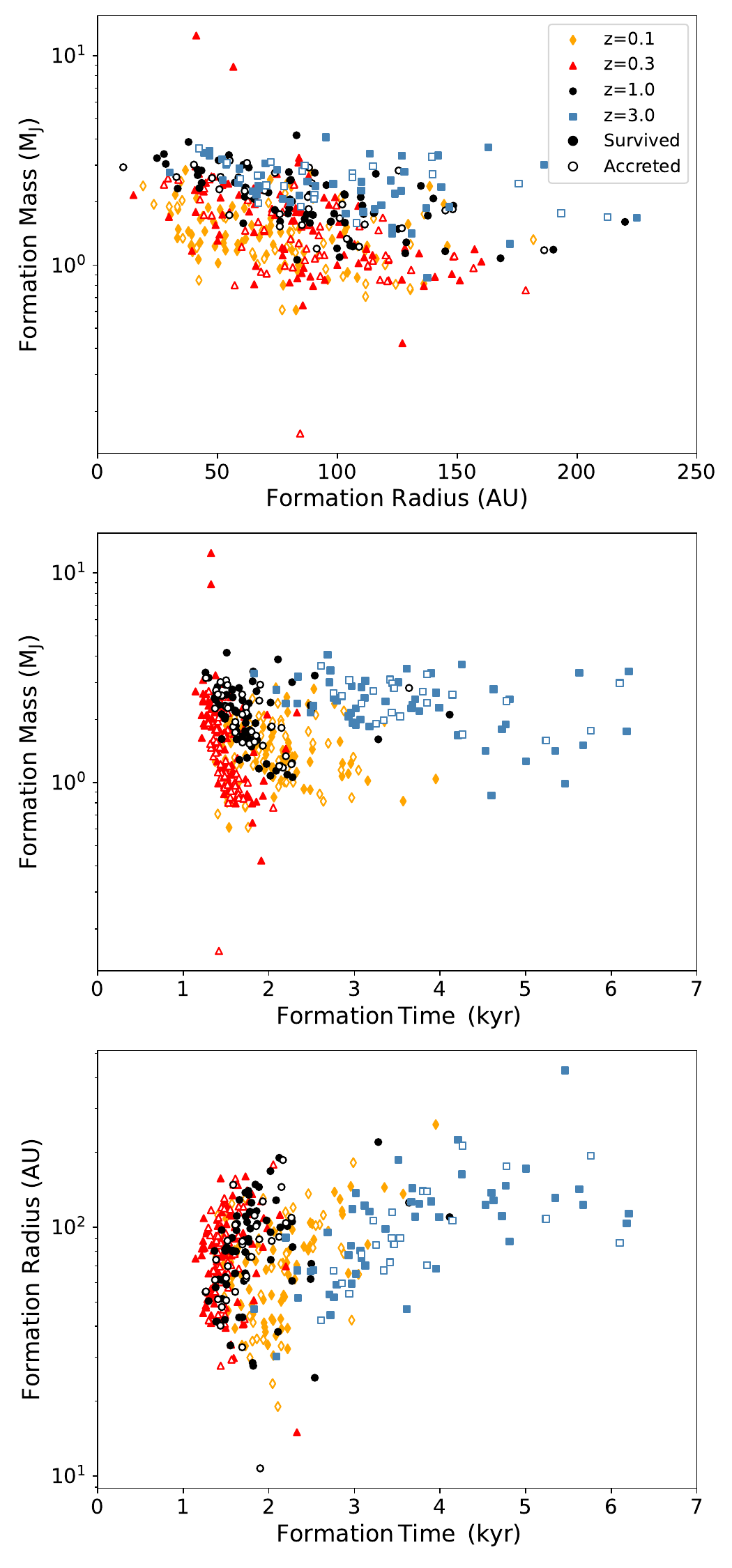}
    \caption{The formation mass, orbital radius and time at of the protoplanets formed in the simulations of different metallicity. Protoplanets that survive until the end of the simulations (i.e, after the disc has lost 70\% of its original mass) are plotted using solid points, whilst protoplanets that have been merged with the host star or another protoplanet are plotted as hollow points. From top to bottom, we show: (a) the formation mass against formation orbital radius of protoplanets, (b) the formation time against formation mass, and (c) the formation orbital radius formation time against formation time.}
    \label{fig:initial_scatters}
\end{figure}

\begin{figure}
    \centering
    \includegraphics[width=0.9\columnwidth]{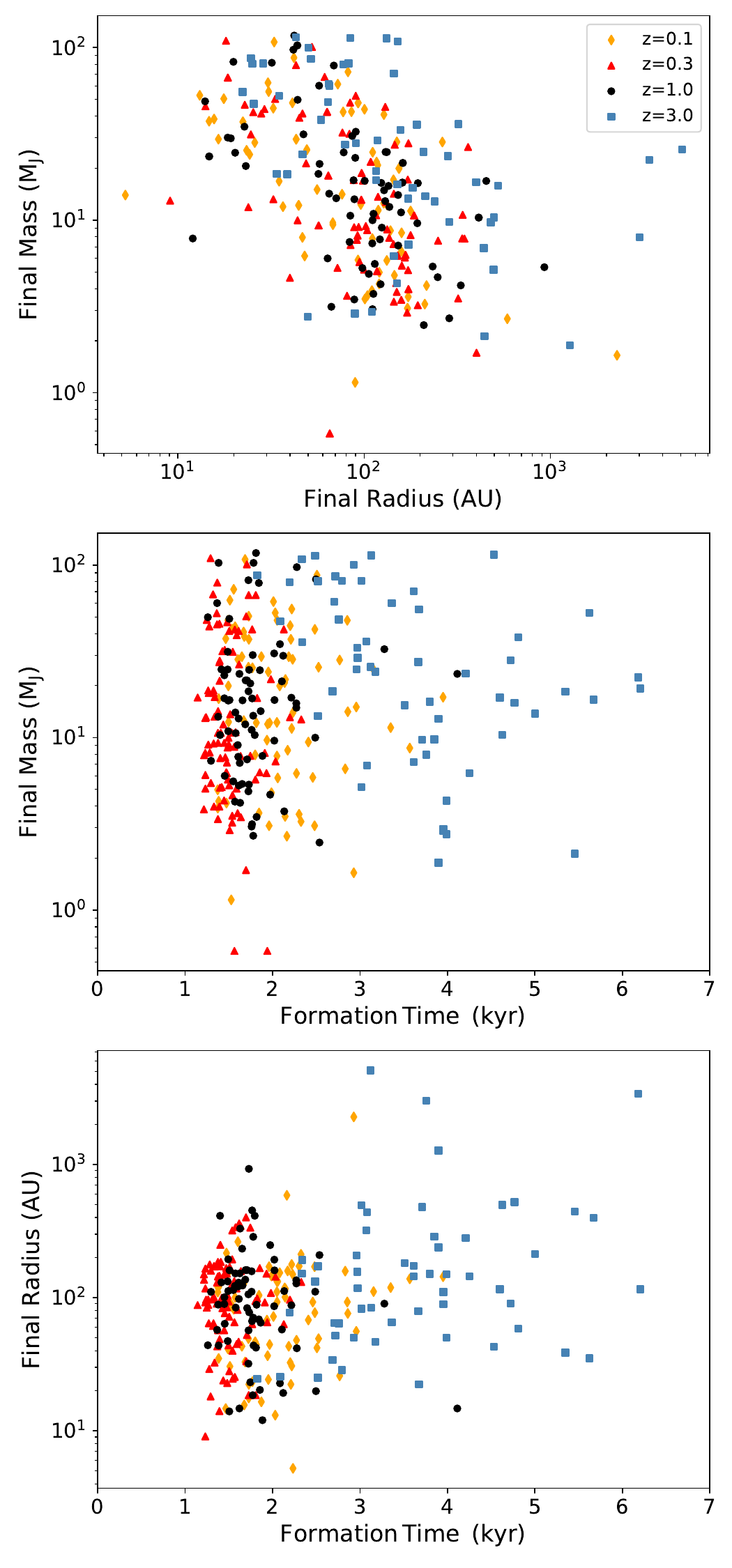}
    \caption{The final mass, orbital radius and formation time at of the protoplanets that survive until the end of the simulations, for each metallicity simulated. From top to bottom, we show: (a) the final separation against final mass for the protoplanets, (b) the formation time against final mass, and (c) the final orbital radius against formation time.}
    \label{fig:final_scatters}
\end{figure}

The formation mass, orbital radius and formation time of the protoplanets  in the simulations with different metallicities are shown in Fig.~\ref{fig:initial_scatters}, and their values at the end of the simulations are shown in Fig.~\ref{fig:final_scatters}.
The formation time (i.e. how fast the disc fragments) is minimum for discs with $ Z=0.3\,\rm{Z_{\odot}}$, where the cooling rate is maximum.
Fig.~\ref{fig:initial_scatters}a shows that protoplanets that form in discs with higher metallicity are typically more massive, and form at larger orbital separations from their host star.  As the system evolves, the protoplanets  interact gravitationally with each other and the central star, leading to planet-planet scattering events and mergers. Additionally, they interact with the gas in the disc leading to mass growth and migration. These processes modify the imprint of the  initial properties by the end of the simulation, but some trends are still seen, and are discussed in the following subsections. 

\subsubsection{Protoplanet formation times}

\begin{figure}
    \centering  \includegraphics[width=0.95\columnwidth]{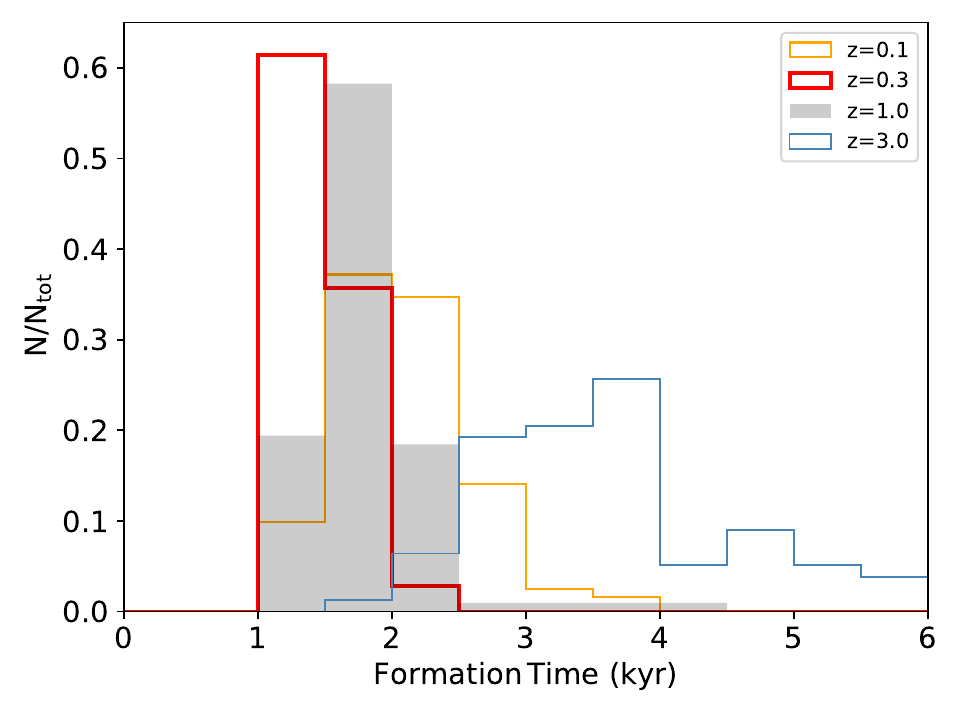}
    \caption{Distribution of the formation times of protoplanets, for discs with different metallicities. Discs with $ Z=0.3\,\rm{Z_{\odot}}$ typically fragment and form protoplanets quicker, whilst discs with $ Z=0.1, 3\,\rm{Z_{\odot}}$ tend to fragment later.}
    \label{fig:formation_time_hist}
\end{figure}

In our simulations, discs fragment within a few kyr; overall, discs with $ Z=0.3,1\,\rm{Z_{\odot}}$, fragment faster (within $1-2$~kyr) than discs with metallicities $ Z=0.1,3\,\rm{Z_{\odot}}$ ($>1.5$~kyr; see Figs. \ref{fig:disc_evolution}, \ref{fig:formation_time_hist}).  Fragmentation occurs the earliest and most rigorously in discs with $ Z=0.3\,\rm{Z_{\odot}}$, (Kolmogorov-Smirnov tests (KS-test hereafter) yield a value of ${p}<10^{-4}$ when comparing  to the distributions of all other metallicities). This behaviour can be attributed to more efficient cooling in discs with $Z = 0.3\,\rm{Z_{\odot}}$ (see Discussion), resulting in shorter cooling timescales. The high-metallicity disc ($Z = 3.0\,\rm{Z_{\odot}}$) cools less efficiently because it is more optically thick than discs with $ Z=0.3\,\rm{Z_{\odot}}$, whereas the low-metallicity disc ($Z = 0.1\,\rm{Z}_{\odot}$) 
cools less efficiently because it is more optically thin.

\subsubsection{Protoplanet mass distributions}

\begin{figure}
    \centering
    \includegraphics[width=0.95\columnwidth]{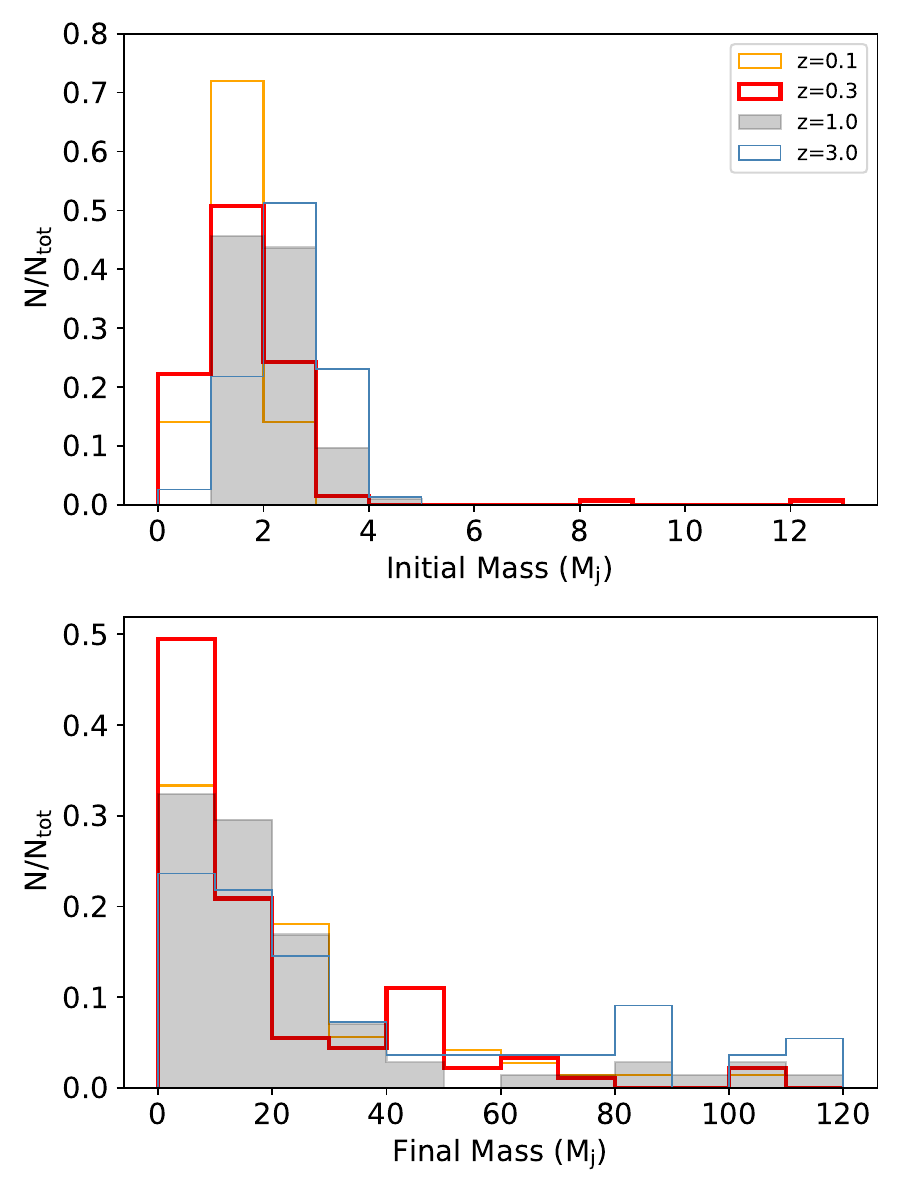}
    \caption{The distribution of masses for all protoplanets formed in the simulations, for different metallicities, as marked on the graph. We show the formation masses (a, top), and final masses,
 i.e. after the discs have lost $70\%$ of their mass (b, bottom).}
    \label{fig:mass_hists}
\end{figure}

Fig. \ref{fig:mass_hists} presents the distribution of the initial and final masses of the protoplanets formed in the simulations. Protoplanets initially form with masses on the order of a few $\rm M_{\rm J}$ (Fig.~\ref{fig:mass_hists}a), which is as expected from the opacity limit for gas fragmentation \citep{Rees:1976a, Whitworth:2006a, Stamatellos:2009b}. Simulations with lower disc metallicities ($Z=0.1\,\rm{Z_{\odot}}$) tend to produce more lower mass protoplanets, with typical masses $\leq2\rm{M_{J}}$.
This is because the minimum mass of fragmentation (i.e. the opacity mass limit) scales weakly with the opacity/metallicity,  $M_{\rm min}\propto \kappa^{1/3} T^{5/6}$\citep{Whitworth:2006a}.
In order to test whether the $Z=0.1\,\rm{Z_{\odot}}$ distribution is indeed different from the rest, we perform a KS-test; these yield p-values of ${p}<0.04$, confirming that these distributions are drawn from different populations. On the other hand, we find no strong correlation between metallicity and the final protoplanet mass (see Fig.~\ref{fig:mass_hists}b), which shows that the final protoplanet masses are determined by evolution within the protostellar disc (gas accretion, interactions with other protoplanets) rather than birth. We do note that around $50\%$ of protoplanets formed in discs with $ Z=0.3\,\rm{Z_{\odot}}$ have final masses $\leq 10\rm{M_{J}}$ (KS-tests yield values ${p}<0.02$ when compared to the distributions of the other metallicities). This is possibly due to the fact that cooling is most efficient in these discs (see Discussion), leading to the highest mean number of protoplanets formed per disc (see Table~\ref{tab:stats}). As a result the gas in the disc is distributed over a larger number of protoplanets, leading to more lower-mass protoplanets. 

\subsubsection{Protoplanet orbital radii distributions}

\begin{figure}
    \centering
    \includegraphics[width=0.95\columnwidth]{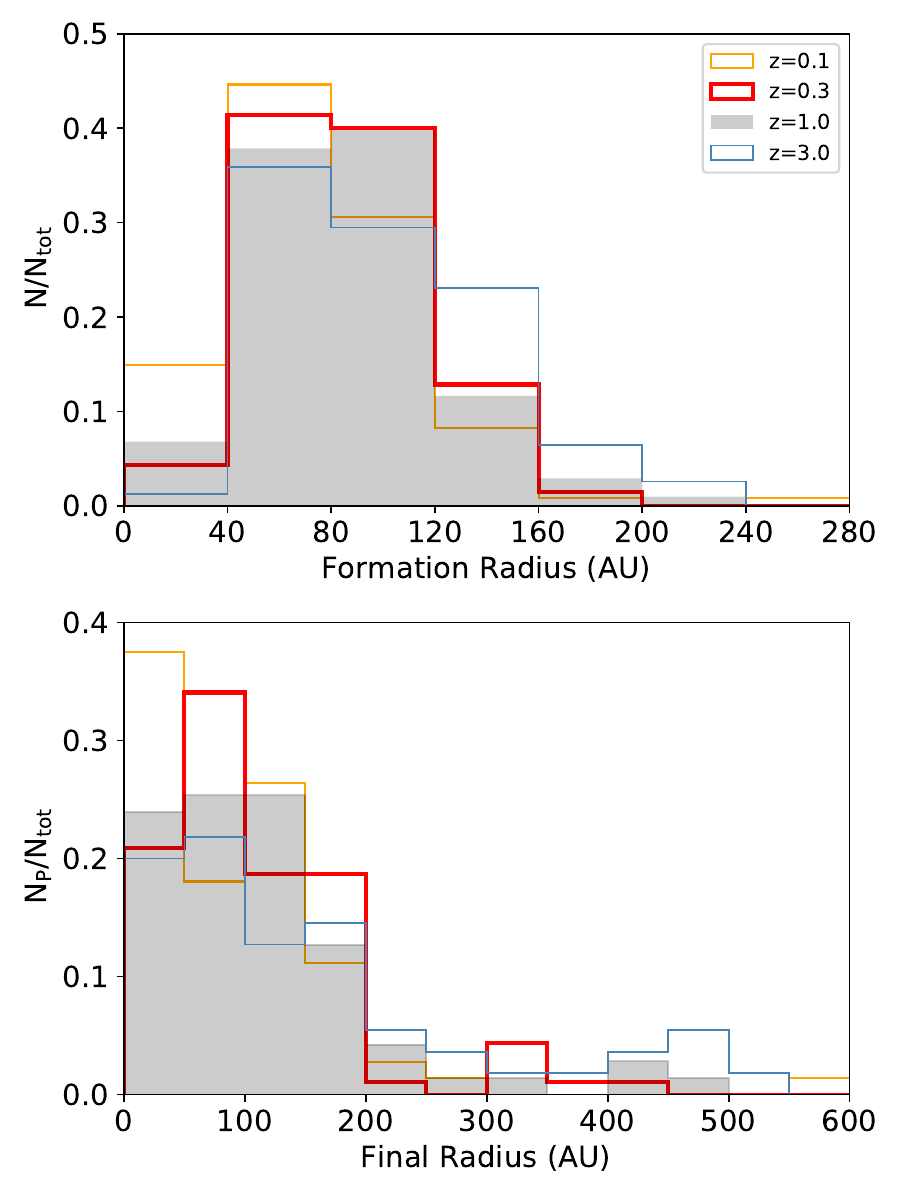}
    \caption{The distribution of orbital radii for all protoplanets formed in the simulations, for different metallicities. We show the orbital radius at formation (a, top), and the final orbital radius  (b, bottom).}
    \label{fig:sep_hists}
\end{figure}

The distributions of the initial and final orbital radii of the protoplanets, for different metallicities are seen in Fig.~\ref{fig:sep_hists}. Discs of all metallicities are more likely to form planets at $40-120$\,AU, with only slight differences among them. In discs with the lowest metallicity at which fragmentation occurs ($Z = 0.1\,\rm{Z}_{\odot}$), more protoplanets form at smaller orbital radii ($\leq 80\,\mathrm{AU}$) than in other models (see Fig. \ref{fig:sep_hists}a; KS-tests yield ${p}<0.04$ when compared to simulations with other metallicities). On the other hand, in higher metallicity discs (e.g. $Z = 3\,\rm{Z}_{\odot}$), there are more protoplanets forming at larger radii ($>120$\,AU) than in other metallicity discs. This is likely linked to the dependence of disc cooling on metallicity as a function of radius. Gas cooling is most efficient when $\tau \sim 1$; lower-metallicity discs satisfy this condition at higher surface densities than higher-metallicity discs, and therefore at smaller radii, i.e. closer to the central star. A similar trend is observed in the distribution of the final radii, suggesting that inward migration is more efficient in lower-metallicity discs. Across all metallicities a small number of protoplanets are also scattered outwards to large orbital radii (> 300\,AU). Therefore, wide-orbit planets may initially form close to the star and migrate outwards as the system evolves dynamically.

\begin{figure}
    \centering
    \includegraphics[width=0.95\columnwidth]{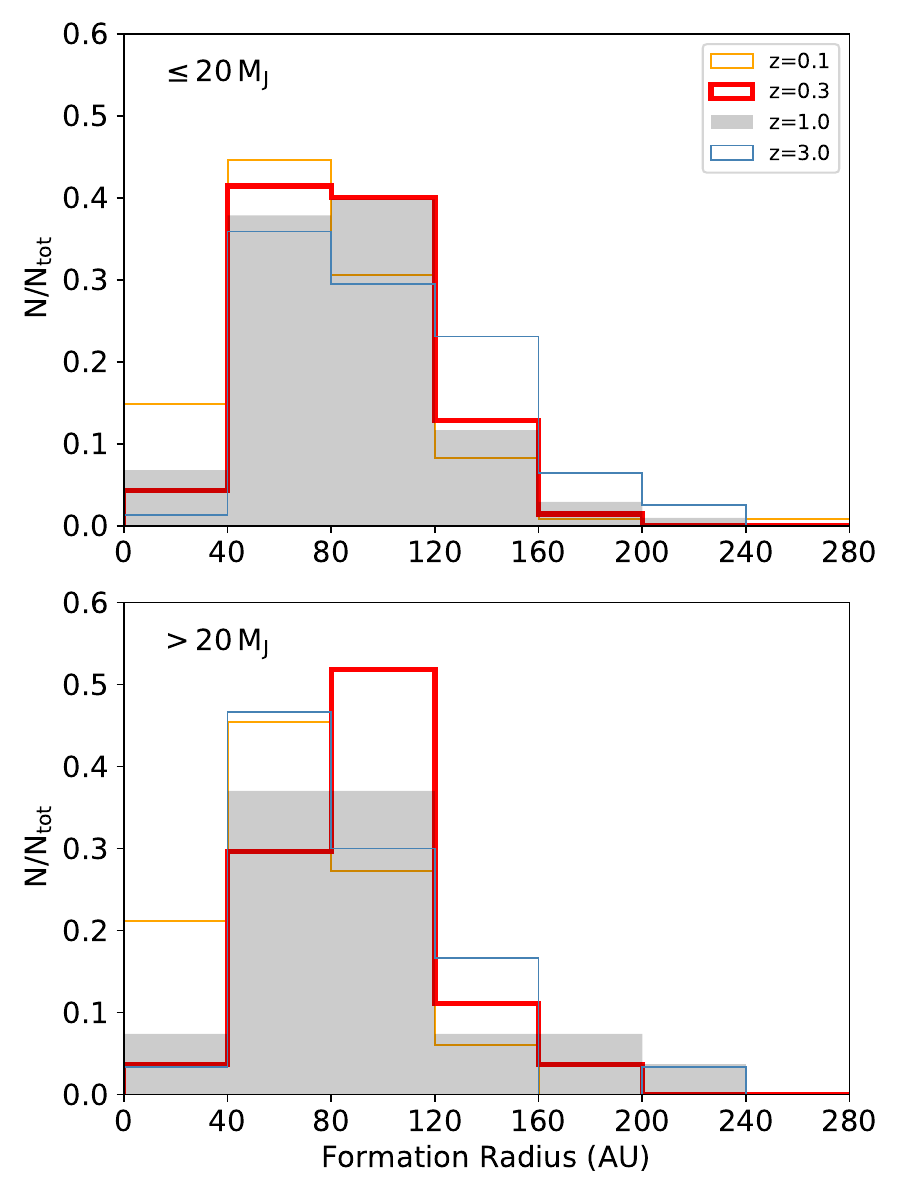}
    \caption{The distribution of  the formation radii of protoplanets that survive until the end of the simulation, split according to mass: (a, top)  protoplanets with final masses $\leq20\rm\,M_{J}$, which will likely evolve into planets, and (b, bottom)  protoplanets with final masses $>20\rm\,M_{J}$, which may evolve into brown dwarfs.}
    \label{fig:jups_BDs_formation_radius}
\end{figure}

Fig.~\ref{fig:jups_BDs_formation_radius} shows the distribution of the formation radii of the protoplanets, separated between those with final masses above and below $20\,\rm{M_{J}}$. Both sets of distributions are similar, with only small differences between them. More protoplanets with final masses $\leq20\,\rm{M_{J}}$ form at larger orbital radii in discs with the highest metallicity than in discs with lower metallicities (see Fig. \ref{fig:jups_BDs_formation_radius}a, KS-tests yields ${p}<0.02$ when compared to the distributions of other metallicities). There is no strong correlation between formation radius and metallicity for protoplanets with final masses $>20\,\rm{M_{J}}$.

\begin{figure}
    \centering
    \includegraphics[width=0.95\columnwidth]{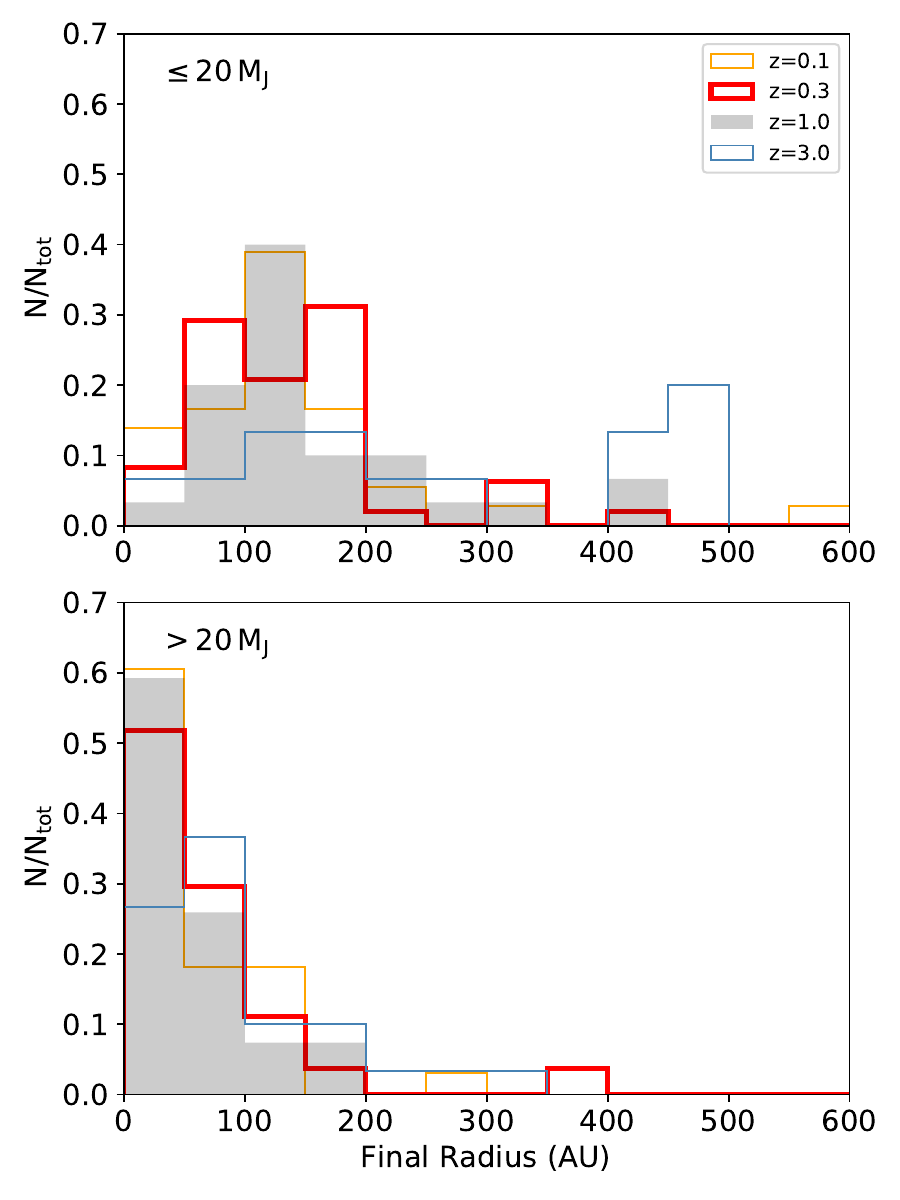}
    \caption{The distribution of the final radii of protoplanets that survive until the end of the simulation, split according to mass: (a, top)  protoplanets with final masses $\leq20\rm\,M_{J}$, and (b, bottom)  protoplanets with final masses $>20\rm\,M_{J}$.}
    \label{fig:jups_BDs_final_radius}
\end{figure}

 The distribution of the final radii of the protoplanets, separated by mass range are seen in Fig.~\ref{fig:jups_BDs_final_radius}.
The less massive protoplanets (${M_{\rm p}}\leq20\rm\,M_{J}$, see Fig. \ref{fig:jups_BDs_final_radius}a) have a broader distribution of final orbital radii, peaking at at $100-150$\,AU whilst the more massive protoplanets ($>20\rm\,M_{J}$)  are typically found within $100\rm\,AU$ of their host star (see Fig.~\ref{fig:jups_BDs_final_radius}b). This is because protoplanets accrete more gas in the inner disc regions, thereby growing in mass, while lower-mass protoplanets are dynamically scattered outwards from this region \citep{Stamatellos:2009b, Teasdale:2026b}.
Only minor differences are found among the distributions for different metallicities, with the exception of the highest-metallicity discs ($Z = 3\,\rm{Z}_{\odot}$), which tend to have fewer low- and high-mass protoplanets within 100\,AU of the host star compared to discs with other metallicities.

\subsection{Protoplanet accretion}

\begin{figure}
    \centering
    \includegraphics[width=0.95\columnwidth]{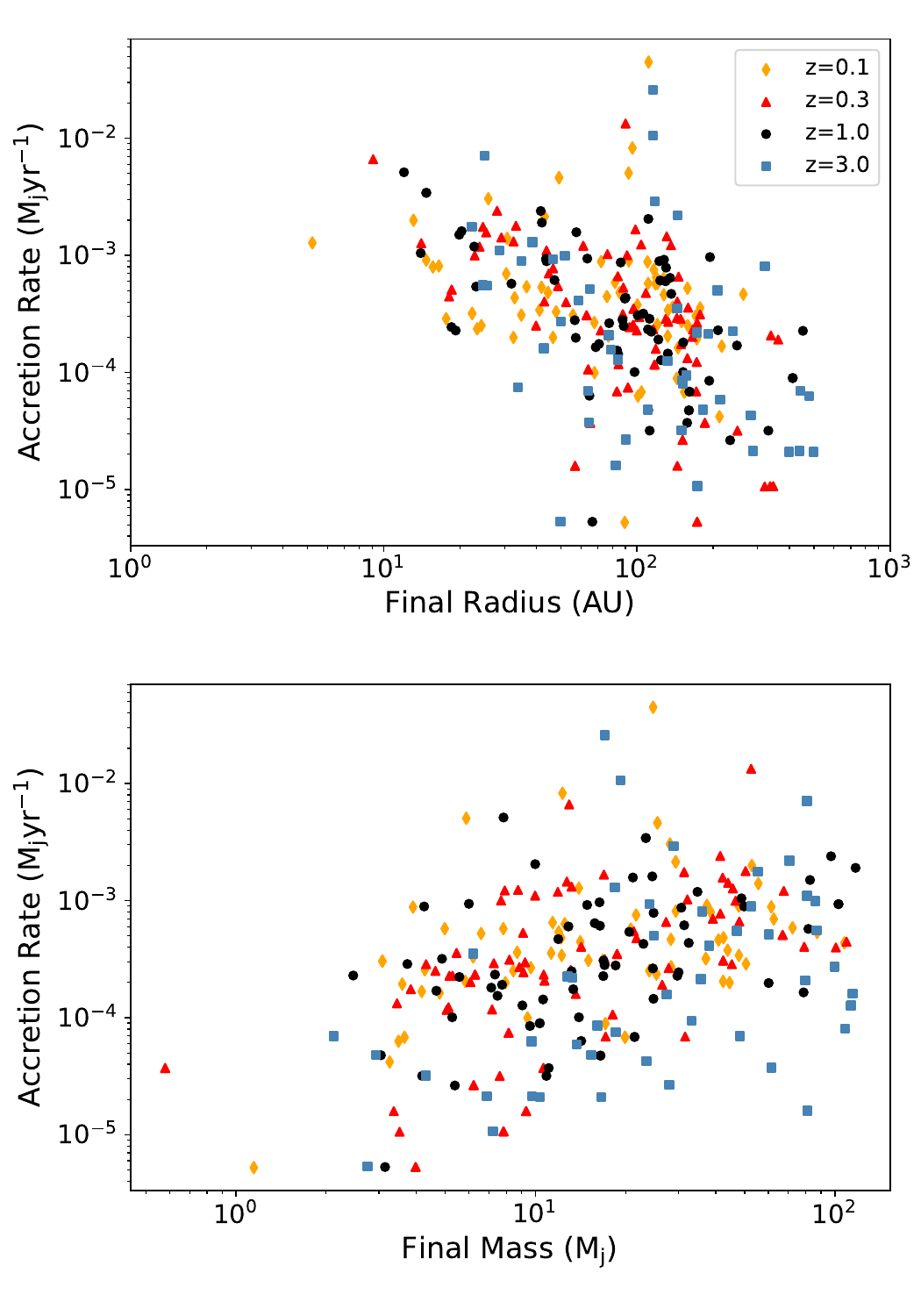}
    \caption{The final accretion rate of  protoplanets with respect to their final orbital radius (a, top) and mass (b, bottom).}
    \label{fig:final_vs_accretion}
\end{figure}

Protoplanets remain actively accreting at the end of the simulations. Fig.~\ref{fig:final_vs_accretion} presents the final accretion rate onto the protoplanets as a function of orbital radius (a), and mass (b). The accretion rates span several orders of magnitude, from $10^{-5}$ to $10^{-2}\,{\rm M_J\,yr^{-1}}$. We find that the accretion rate generally decreases with increasing orbital radius and increases with protoplanet mass.

The estimated accretion rates are up to two orders of magnitude higher than those estimated for the embedded planets PDS~70~b,c \citep{Keppler:2018a,Haffert:2019d}, which accrete at $\sim10^{-8}$--$10^{-7}\,{\rm M_J\,yr^{-1}}$ \citep{Wagner:2018a,Wang:2020a,Zhou:2021a}. AB~Aurigae~b \citep{Currie:2022q,Currie:2025a}, another disc-embedded planet, accretes at a higher rate of $\sim10^{-6}\,{\rm M_J\,yr^{-1}}$ \citep{Currie:2022q,Zhou:2022a}, closer to the values found here. In contrast, WISPIT~2b \citep{Li:2025,van-Capelleveen:2025a,Close:2025a}, a wide-orbit ($\sim60$\,AU) gap-embedded planet, exhibits a much lower accretion rate of $\sim10^{-12}\,{\rm M_J\,yr^{-1}}$ \citep{Li:2025}.

\begin{figure}
    \centering
    \includegraphics[width=0.95\columnwidth]{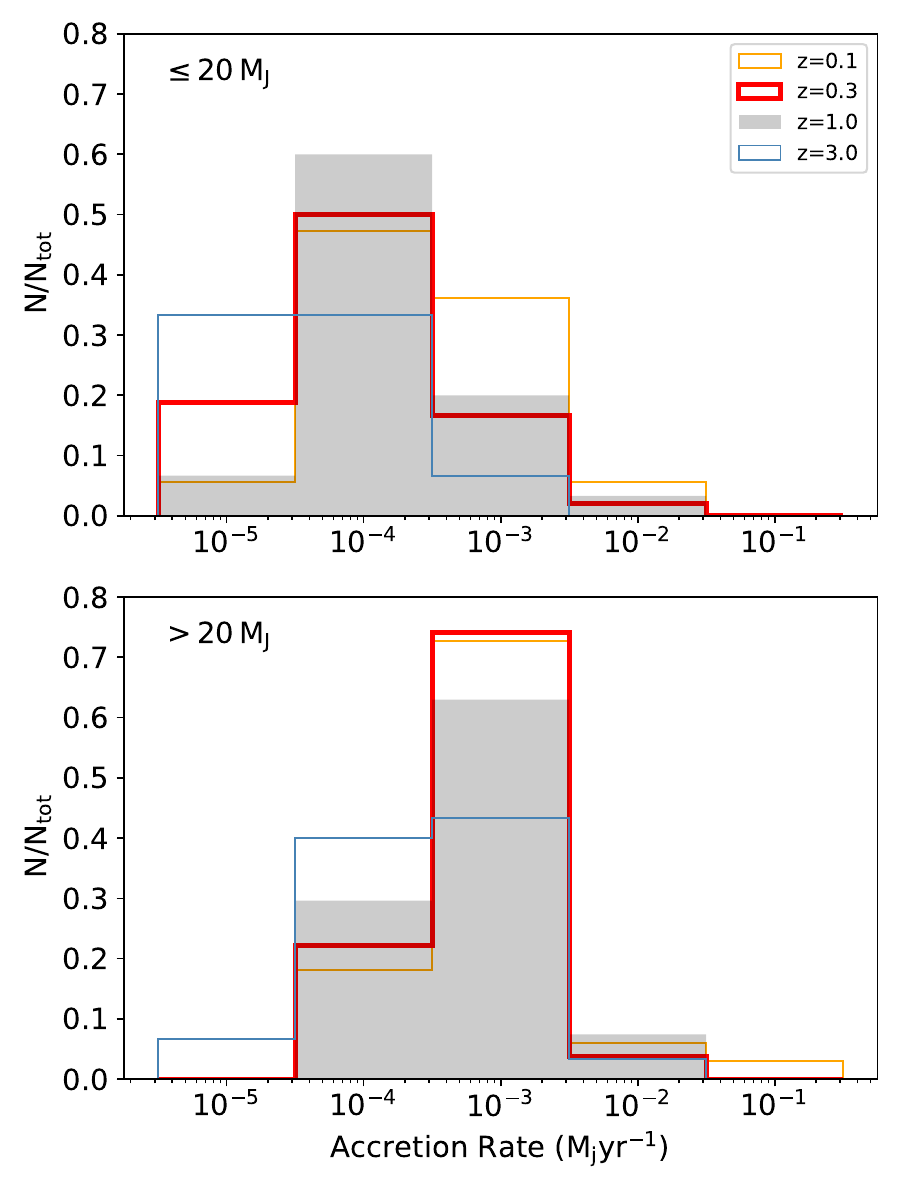}
    \caption{The distribution of the  final accretion rates of protoplanets, split according to protoplanet mass: (a, top)  protoplanets with final masses $\leq~20\rm\,M_{J}$, and (b, bottom) protoplanets with final masses $>20\rm\,M_{J}$.}
    \label{fig:jups_BDs_final_accretion}
\end{figure}

We find that protoplanets with masses $\leq 20\,\mathrm{M_J}$ are generally accreting at lower rates than those with masses $> 20\,\mathrm{M_J}$ (see Fig.~\ref{fig:jups_BDs_final_accretion}). We do not find a strong dependence of the final accretion rate on metallicity. However, protoplanets in high-metallicity discs tend to accrete at lower rates, likely because less efficient cooling suppresses gas accretion \citep{Nayakshin:2017a, Stamatellos:2018a, Mercer:2018}. It is important to note that sink accretion rates in SPH simulations tend to be overestimated, so the accretion rates presented in Fig. \ref{fig:jups_BDs_final_accretion} may be higher than the real accretion rates expected for protoplanets. As a result, the final protoplanet masses we find in this work provide an upper limit for the mass of objects produced by gravitational instability. 


\subsection{Protoplanet migration}

\begin{figure}
    \centering
    \includegraphics[width=0.95\columnwidth]{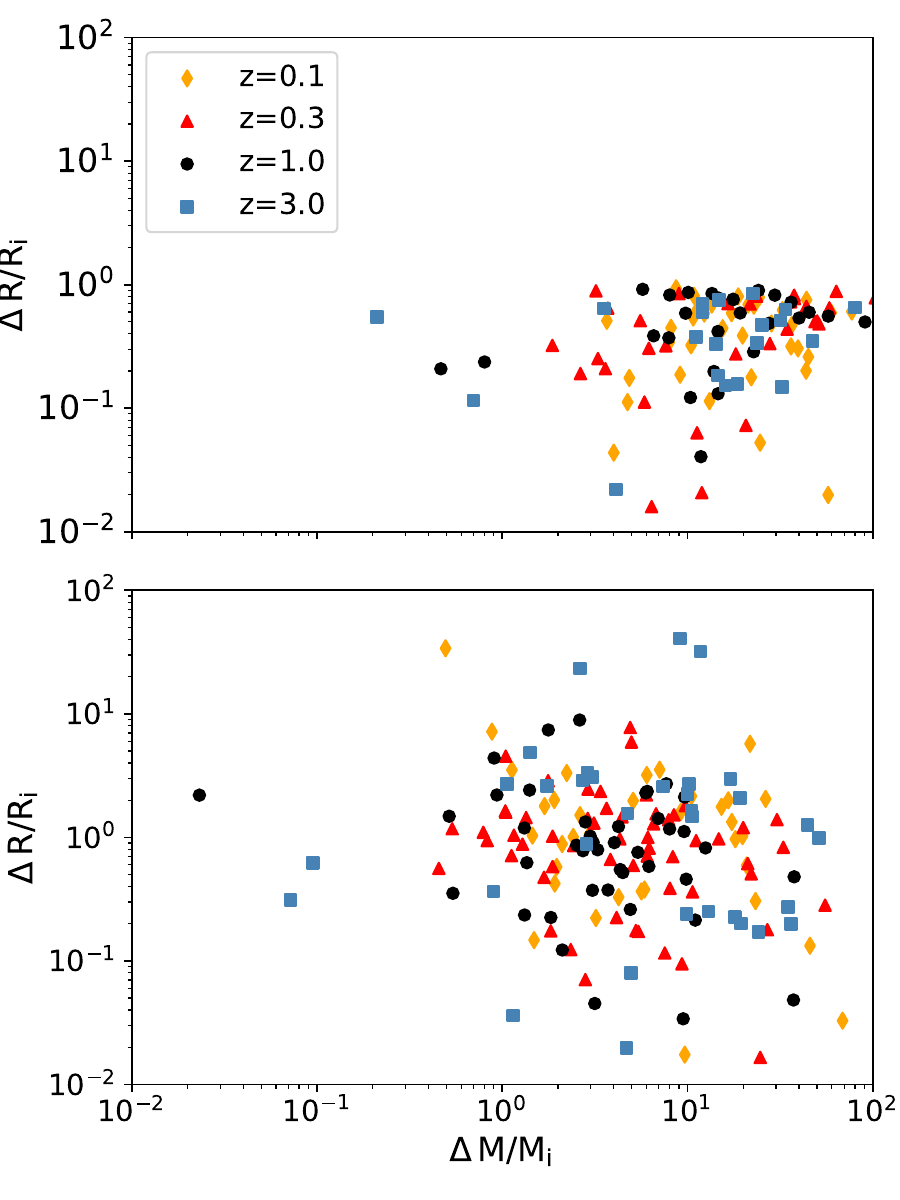}
    \caption{The fractional change of mass against the fractional change of orbital separation for the protoplanets at the end of the simulations: (a, top) protoplanets that experience inwards migration, and (b, bottom) protoplanets that experience outwards migration.}
    \label{fig:dm_dr}
\end{figure}

We use the term {\it migration} to denote changes in a protoplanet's orbital radius resulting either from disc-planet interactions or from planet-planet scattering interactions.
 Protoplanets migrate inwards ($40\pm3$\% of protoplanets that survive until the end of the simulation) or outwards ($60\pm5$\%), while they increase in mass either through accreting material from the disc or through merging with other protoplanets (see Fig. \ref{fig:dm_dr}). 

 We find no correlation between metallicity and changes in mass or orbital radius, which suggests that migration is not primarily driven by disc-planet interactions \citep{Baruteau:2014a, Durmann:2015a, Stamatellos:2015a}, but instead by stochastic dynamical interactions (i.e.\ scattering) with other protoplanets and the central star \citep[e.g.][]{Stamatellos:2009b}.

\subsection{Protoplanet collapse timescale}

Clumps that form in gravitationally unstable discs typically evolve through the first collapse, first core formation , second collapse, and second core formation phases \citep{Stamatellos:2009a}, i.e. in an evolutionary pathway similar to that of a solar-mass collapsing cloud core \citep{Larson:1969,Masunaga:2000,Stamatellos:2007,Bhandare:2018,Young:2023}. 

Initially, the clump undergoes an isothermal, free-fall collapse phase in which
the gravitational energy provided by the collapse is radiated away and does not heat the clump. The density in the center increases  until it reaches a critical value of $\rho\sim10^{-12}\rm\,g\,cm^{-3}$; at this density the clump becomes optically thick, with the temperature  and density rapidly increasing resulting in  hydrostatic equilibrium and the formation of the first hydrostatic core. The collapse then becomes adiabatic with the temperature and density increasing gradually. When the central temperature reaches $\,T\sim2,000$~K, $\rm\,H_{2}$ dissociates. At this stage the clump undergoes a second collapse eventually forming the second hydrostatic core (i.e. the protoplanet). We follow this evolution up to  a density of $\rho=10^{-3}\rm\,g\,cm^{-3}$ \citep{Stamatellos:2009b, Mercer:2020,Fenton:2024}. 

\begin{figure}
    \centering
    \includegraphics[width=0.95\columnwidth]{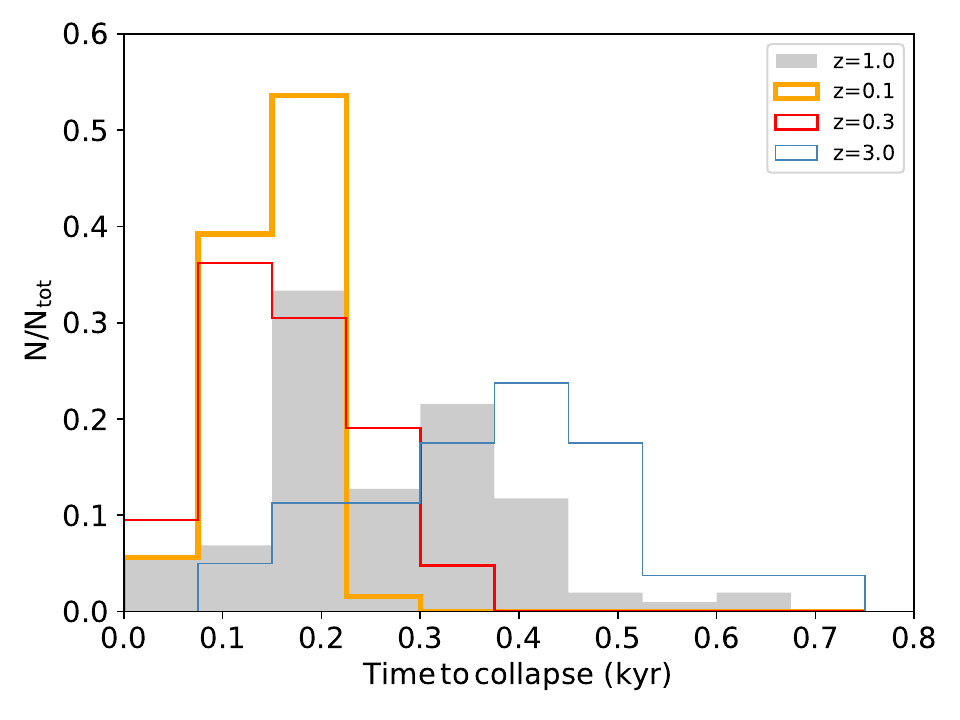}
    \caption{Distribution of the clump collapse time for different metallicities. Clumps that form in discs with $ Z=0.1\,\rm{Z_{\odot}}$ typically collapse faster, as they cool more efficiently.}
    \label{fig:collapse_hist}
    \centering
    \includegraphics[width=0.95\columnwidth]{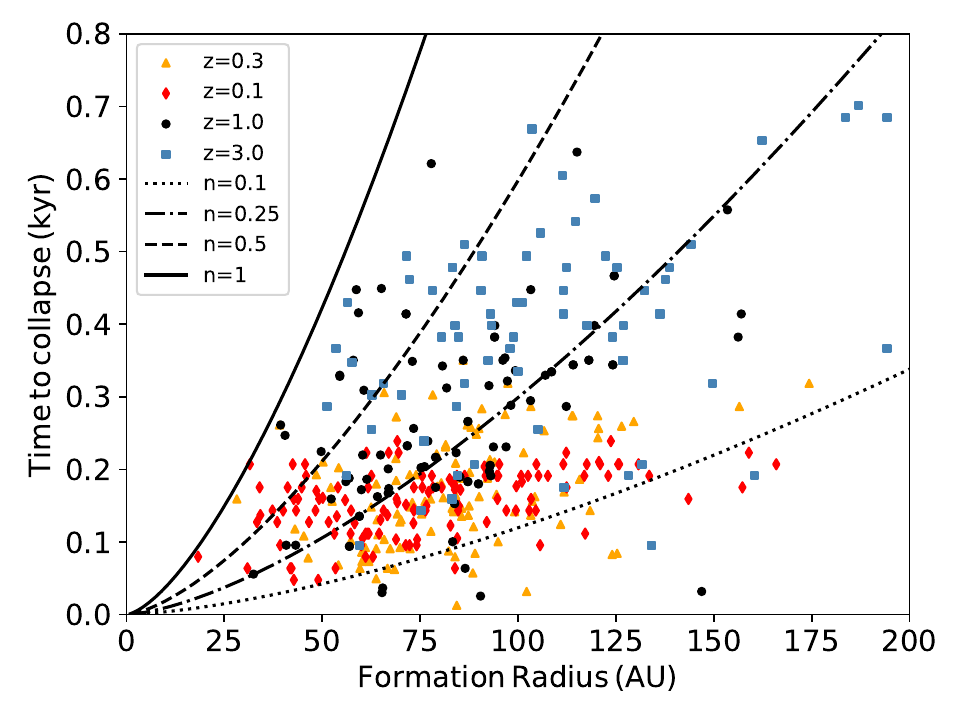}
    \caption{The time that it takes for clumps to collapse from $10^{-3}\rm\,g\,cm^{-3}$ to $10^{-9}\rm\,g\,cm^{-3}$, becoming protoplanets, against their formation radius, for different metallicities. We also include lines for $n=0.1,\,0.25,\,0.5$ {\rm and}\ $1$ orbital periods. All clumps collapse within $1$ orbital timescale.}
    \label{fig:collapse_orbit}
\end{figure}

We define the protoplanet formation timescale as the time required for a clump to collapse from a central density of $\rho_{\rm c} = 10^{-9}\,\mathrm{g\,cm^{-3}}$ to $\rho_{\rm c} = 10^{-3}\,\mathrm{g\,cm^{-3}}$ (see Fig.~\ref{fig:collapse_hist}).
The collapse timescale is on the order of a few hundred years (Fig.~\ref{fig:collapse_orbit}), with clumps forming on wider orbits taking more time to collapse. The collapse timescale is shorter than the local orbital period, leaving only a limited time for clump dispersal via tidal interactions with other clumps or spiral arms in the disc. We find that the collapse timescale increases with metallicity (Fig.~\ref{fig:collapse_orbit}). At these high densities, the clumps are optically thick (even at low metallicities); consequently, the cooling timescale (see Eq.~\ref{eq:optical_depth}) increases with optical depth, and therefore with metallicity.

\section{Discussion}

We find that disc fragmentation is sensitive  to metallicity but the relation is not monotonic (see Fig. \ref{fig:disc_evolution}). Fragmentation is promoted when the disc cools more efficiently. We find a sweet spot for fragmentation at ${Z}\sim 0.3\,\rm{Z_{\odot}}$ ($-0.5 \,\rm{dex}$); at higher metallicities the disc cools less efficiently because it is more optically thick, whereas at lower metallicities the disc cools less efficiently because it is more optically thin. In the optically thick regime, photons are repeatedly absorbed and re-emitted before escaping the gas. Therefore, radiation propagates through the gas via diffusion (i.e. a random walk process) and the cooling rate decreases as the optical depth increases as photons will take longer to escape. Conversely, in the optically thin regime photons can escape freely without being reabsorbed and the cooling rate is determined by the local emissivity of the gas. The cooling rate becomes less efficient, and is insufficient to overcome heating from the central star and from PdV work within the disc (i.e. shock heating in the spiral arms). As a result disc fragmentation is suppressed in both extremely metal-rich discs (${Z}\gtrsim 3\,\rm{Z_{\odot}}$) and extremely metal-poor discs (${Z}\lesssim 0.1\,\rm{Z_{\odot}}$). 

Cooling is more efficient when the optical depth is $\tau=z\kappa \Sigma \sim1$ (see Eq.~\ref{eq:optical_depth}), where $\kappa$ is the opacity for $Z=1\,\rm{Z_{\odot}}$, and $z$ sets the metallicity. A typical orbital radius for disc fragmentation is $\sim 80$\,AU (see Fig.~\ref{fig:sep_hists}), where the disc surface density is $\Sigma\sim 40\ {\rm g cm^{-2}}$ (see Eq.~\ref{eq:sigma}), the temperature $T\sim20$~K (see Eq.~\ref{eq:temp}), and the corresponding opacity $\kappa\sim 0.08\ {\rm cm^{2} g^{-1}}$ (see Table~\ref{tab:opacities}). For these values $\tau\sim 1$ corresponds to $z\sim 0.3$, i.e. $Z=0.3\,\rm{Z_{\odot}}$, which is where we find that fragmentation is more vigorous. 

Discs with metallicities $ Z \gtrsim 0.3\,\rm{Z}_{\odot}$ are optically thick at this radius, whereas discs with metallicities $ Z \lesssim 0.1\,\rm{Z}_{\odot}$ are optically thin; in both cases, the cooling efficiency is reduced (see Eq.~\ref{eq:dudt_rad}). However, there is a range of surface densities and temperatures across the disc, corresponding to different optical depths. Consequently, fragmentation may be promoted at different radii depending on the disc metallicity (see Fig.~\ref{fig:sep_hists}). 

An exception arises when the metallicity is sufficiently low that the entire disc becomes too optically thin ($ Z \lesssim 0.1\,\rm{Z}_{\odot}$), reducing the cooling efficiency. Similarly, at very high metallicities ($ Z \gtrsim 3\,\rm{Z}_{\odot}$), the disc becomes too optically thick throughout, again reducing the cooling efficiency and fully suppressing fragmentation.
 
Our results disagree with \cite{Boss:2002}, who find that disc fragmentation shows no dependence on metallicity, but are in agreement with \cite{Cai:2006}, who find that GI becomes stronger as the metallicity decreases from $2\,\rm Z_{\odot}$ down to $0.25\,\rm Z_{\odot}$. 
We also find agreement with \cite{Lee:2025}, who investigate the effect of dust opacity on disc fragmentation. They vary the opacity as a function of temperature and grain size to explore the role of dust growth in planet formation and disc fragmentation. This is effectively analogous to our approach: they attribute opacity variations to dust growth, whereas we relate them to metallicity, i.e. the dust abundance. They similarly find that fragmentation is promoted with decreasing opacity, consistent with the trend we observe from $3\,\rm Z_{\odot}$ down to $0.3\,\rm Z_{\odot}$.  \cite{Meru:2010}, who study fragmentation of  discs with opacities ranging from $0.01$ to $10$ times the interstellar values, find that fragmentation is promoted in lower opacity discs due to more efficient cooling. However, they do not find a sweet spot for fragmentation; this is possibly due to adopting the $\beta$-cooling approximation instead of a more detailed description of the disc thermal physics.

\cite{Matsukoba:2022,Matsukoba:2023} find that fragmentation can occur at lower metallicities ($10^{-6}\,\rm Z_{\odot}$ to $1\,\rm Z_{\odot}$). However, in contrast to our results, they find that fragmentation remains possible even at extremely low metallicities, even at zero metallicity. They report that fragmentation is more vigorous in the range $10^{-5}\,\rm Z_{\odot}$ to $10^{-2}\,\rm Z_{\odot}$, and more modest for higher metallicities in the range $0.1\,\rm Z_{\odot}$ to $1\,\rm Z_{\odot}$. However, their models include additional cooling processes that may become significant at low metallicities, such as molecular (e.g. H$_2$, HD) and fine-structure line emission.

Our work may explain the slight overabundance of wide-orbit giant planets observed around metal-poor stars (see Fig.~\ref{fig:obs_metallicity_hists_jups}); these are likely to have formed via disc fragmentation. Since discs at metallicities $Z = 0.3\,\rm{Z}_{\odot}$ fragment more vigorously, this may imply that giant planet formation via disc instability is more efficient in environments with sub-solar metallicities. In particular, regions further out in the Galaxy, where the typical metallicity is lower \citep{Yasui:2026a}, or metal-poor galaxies may host a higher incidence of such planets. However, this interpretation remains uncertain, as several factors may influence disc fragmentation in these environments, including the formation and initial conditions of discs within molecular clouds of different metallicities. 

Furthermore, we neglect the effects of dust evolution in this work. Previous studies have shown that the distribution of dust in the disc is inhomogeneous \citep{Booth:2016, Longarini:2023, Birnstiel:2024, Rowther:2024}. Dust is preferentially concentrated in the spiral arms, which may increase the opacity, thereby affecting gas cooling  and disc fragmentation. Further study of the effects of dust evolution on the local opacity in gravitationally unstable discs alongside proper treatment of the thermodynamics is required to fully understand the relationship between metallicity and disc fragmentation. 

\section{Conclusions}

We used the SPH code \textsc{phantom} to study the gravitational fragmentation of massive circumstellar discs with varying metallicity. We performed multiple simulations of a disc of mass $0.25\,\rm M_{\odot}$ orbiting a $0.7\,\rm M_{\odot}$ star. The disc metallicity is varied from $0.01$ to $10\,\rm{Z}_{\odot}$ in order to investigate its effect on the evolution of gravitationally unstable protoplanetary discs and on the properties of protoplanets formed if these discs fragment.

We find that disc fragmentation is sensitive to metallicity, with a sweet spot at $Z = 0.3\,\rm{Z}_{\odot}$ where fragmentation is fastest and most vigorous, producing the largest number of protoplanets per disc. Fragmentation still occurs at lower (down to $0.1\,\rm{Z}_{\odot}$) and higher (up to $3\,\rm{Z}_{\odot}$) metallicities. However, it is suppressed at extremely low and high metallicities ($0.01\,\rm{Z}_{\odot}$ and $10\,\rm{Z}_{\odot}$, respectively), as such discs cannot cool efficiently.

The initial masses and orbital radii of protoplanets formed in the simulations show a weak dependence on metallicity. Lower-metallicity protoplanets tend to have higher initial masses and form closer to the star than their higher-metallicity counterparts. After evolving within the disc, this trend persists, with a larger fraction of low-metallicity protoplanets remaining at lower masses and smaller orbital radii.
Low-metallicity protoplanets also collapse faster than higher-metallicity ones, allowing only a limited time for disruption by tidal interactions.

Disc metallicity has a significant impact on disc evolution, determining the onset and strength of fragmentation as well as the initial properties of protoplanets. We conclude that GI is a plausible mechanism for forming gas giant planets and brown dwarfs in discs spanning a range of metallicities ($0.1\,\rm{Z}_{\odot} \leq Z \leq 3\,\rm{Z}_{\odot}$), with peak efficiency at the sub-solar metallicity of $0.3\,\rm{Z}_{\odot}$.

\section*{Acknowledgements}
The authors acknowledge support from the COST Action CA22133:PLANETS. EC acknowledges support from STFC grant ST/X508329/1 and DS from STFC grant ST/Y002741/1. The simulations were performed using the University of Lancashire High Performance Computing (HPC) and High Throughput Computing (HTC) facilities, and the Cambridge Service for Data Driven Discovery (CSD3) operated by the University of Cambridge Research Computing Service (www.csd3.cam.ac.uk), provided by Dell EMC and Intel using Tier-2 funding from the Engineering and Physical Sciences Research Council (capital grant EP/T022159/1), and DiRAC funding from the Science and Technology Facilities Council (www.dirac.ac.uk).
The plotting software packages {\sc splash} \citep{Price:2018} and {\sc sarracen} were used \citep{Harris:2023} for parts of the analysis.

\section*{Data Availability}
The simulation data used for this paper can be provided by contacting the authors.



\bibliographystyle{mnras}
\bibliography{refs}

\appendix

\bsp	
\label{lastpage}
\end{document}